\documentclass[]{aastex631}

\usepackage{amsmath}
\usepackage{graphicx}
\usepackage{txfonts}
\usepackage{comment}

\usepackage{color}
\usepackage{listings}
\usepackage{array}
\usepackage{subfigure}
\usepackage{hyperref}
\usepackage{comment}
\usepackage{soul}

\newcommand{\GravC}{\mathcal G} 
\newcommand{\redmass}{\mathfrak{m}}
\newcommand{\GM}{\mu} 

\newcommand{\disk}{\mathrm{disk}}
\newcommand{\pln}{\mathrm{pl}}
\newcommand{\mig}{\mathrm{m}}
\newcommand{\tot}{\mathrm{tot}}
\newcommand{\Lind}{\mathrm{L}}
\newcommand{\Corot}{\mathrm{C}}

\newcommand{\bfr}{\boldsymbol{r}}
\newcommand{\bfv}{\boldsymbol{v}}
\newcommand{\bfa}{\boldsymbol{a}}
\newcommand{\bfk}{\boldsymbol{k}}
\newcommand{\bfF}{\boldsymbol{F}}
\newcommand{\bfe}{\boldsymbol{e}}
\newcommand{\ANGMOM}{\mathcal L}
\newcommand{\bfANGMOM}{\boldsymbol\ANGMOM} 

\newcommand{\h}{H}

\newcommand{\OMEGONA}{\Omega}

\newcommand{\GAMMONA}{\Gamma}

\newcommand{\D}[1]{\ensuremath{\operatorname{d}\!{#1}}} 

\newcommand{\gas}{\mathrm{gas}} 
\newcommand{\accr}{\mathrm{accr}} 
\newcommand{\Kepl}{\mathrm{K}} 
\newcommand{\alphaSigma}{\alpha_\Sigma} 
\newcommand{\betaT}{\beta_T} 
\newcommand{\betah}{{\beta_\mathrm{f}}} 
\newcommand{\alphaturb}{\alpha_\mathrm{t}} 
\newcommand{\nuvisc}{\nu_\mathrm{t}} 
\newcommand{\mpl}{m_{\mathrm{pl}}}

\newcommand{\I}{i}

\begin{document} 
\font\courier=pcrr at 12pt

\title{A recipe for eccentricity and inclination damping for partial gap opening planets in 3D disks}

\author[0000-0003-3622-8712]{Gabriele Pichierri}
\affiliation{Division of Geological and Planetary Sciences, California Institute of Technology, Pasadena, CA 91125, USA}

\author{Bertram Bitsch}
\affiliation{University College Cork, College Rd, University College, Cork, Ireland}
\affiliation{Max-Planck-Institut für Astronomie, Königstuhl 17, 69117 Heidelberg, Germany}

\author{Elena Lega}
\affiliation{Laboratoire Lagrange, Université Cote d’Azur, Observatoire de la Cote d’Azur, 06304 Nice, France}



\begin{abstract}

{
In a previous paper we showed that, like the migration speed, the eccentricity damping efficiency is modulated linearly by the depth of the partial gap a planet carves in the disk surface density profile, resulting in less efficient $e$-damping compared to the prescription commonly used in population synthesis works. Here, we extend our analysis to 3D, refining our $e$-damping formula and studying how the inclination damping efficiency is also affected. We perform high resolution 3D locally isothermal hydrodynamical simulations of planets with varying masses embedded in disks with varying aspect ratios and viscosities. We extract the gap profile and orbital damping timescales for fixed eccentricities and inclinations up to the disk scale height.  The limit in gap depths below which vortices appear, in the low-viscosity case, happens roughly at the transition between classical type-I and type-II migration regimes. The orbital damping timescales can be described by two linear trends with a break around gap depths $\sim80\%$ and with slopes and intercepts depending on the  eccentricity and inclination. These trends are understood on physical grounds and are reproduced by simple fitting formulas whose error is within the typically uncertainty of type-I torque formulas. Thus, our recipes for the gap depth and orbital damping efficiencies yield a simple description for planet-disk interactions to use in N-body codes in the case of partial gap opening planets that is consistent with high-resolution 3D hydro-simulations. Finally, we show examples of how our novel orbital damping prescription can affect the outcome of population synthesis experiments.}

\end{abstract}



\section{Introduction}

One of the goals of a planet formation model is to predict, for a given system or for the full exoplanet population in a statistical sense, what planetary system we expect to form inside a disk with given physical properties (such as surface density, temperature and thickness profiles, a given level of turbulent viscosity, etc.) orbiting a given star \citep{2008ApJ...673..487I, 2009A&A...501.1139M, 2013A&A...558A.109A,
2017MNRAS.464..428A,
2017MNRAS.470.1750I, 
2018MNRAS.474..886N, 2019A&A...623A..88B, 2019MNRAS.486.5690G, 2021A&A...650A.152I, 2021A&A...656A..69E,2023A&A...679A..42S}.
A fundamental ingredient in such a model is the description of planet-disk interactions. A planet embedded in a disk modifies the disk's structure and evolution, and, in turn, this interaction causes the planet's orbit to modify its size, shape, and orientation around the host star.

The presence of a gap in the disk surface density at the location of the planet's orbit is one of the most evident signatures of planet-disk interactions. The existence of gaps in protoplanetary disks is now observationally well established (e.g.\ the ALMA-based DSHARP survey, \citealt{2018ApJ...869L..41A, 2018ApJ...869L..42H} 
; see also \citealt{2020Natur.586..228S}, and \citealt{2023ASPC..534..423B} for a review). Although gaps cannot be linked directly to gap-carving planets as a general rule (because other physical processes in disks can alone explain these features \citep{2017A&A...600A..75B,2019MNRAS.484..107S,2020A&A...639A..95R,2022MNRAS.516.4660C}, and some putative forming planetary systems would actually be dynamically unstable if to each gap corresponded a planet, e.g.\ \citealt{2023A&A...677A..82T}), the presence of a gap in the gas disk is shown in some cases to correspond to the presence of young forming protoplanets \citep{
2018A&A...617A..44K,2018ApJ...863L...8W,2019NatAs...3..749H,2019NatAs...3.1109P,2022ApJ...928....2I}.

If the mere presence of a dip in the surface density can be an indication of the presence of a planet, the depth of the gap that the planet carves (the ratio between the minimum of the surface density profile around the location of the planet and the surface density for the same unperturbed disk) is a crucial parameter to quantitatively describe the interaction between a planet and its surrounding protoplanetary disk.
On the one hand, when a strong enough perturbation has occurred, the planet can create a pressure bump outside of its orbit which prevents pebbles from drifting inwards \citep{2006A&A...453.1129P,2012A&A...546A..18M,2014A&A...572A..35L,2018A&A...615A.110A,2018A&A...612A..30B,2018ApJ...854..153W}; the planet mass at which this occurs is called the pebble isolation mass. In our Solar System, the fact that Jupiter's core would have stopped the inflow of pebbles from the outer disk (the so-called Jupiter barrier) could explain the observed isotopic dichotomy between non-carbonaceous and carbonaceous meteorites \citep{2017PNAS..114.6712K,2020NatAs...4...32K}. The velocity deviation from a pure Keplerian rotation due to the acceleration of gas in response to the presence of a planet (so-called velocity kinks) can also be used to detect the presence of planets (e.g.\ \citealt{2018ApJ...860L..12T,2018ApJ...860L..13P,2019NatAs...3.1109P,2022ApJ...928....2I,2023ASPC..534..645P}).
On the other hand, the carving of a gap and the establishment of a pressure bump affect in turn the planet's own physical and dynamical evolution. By stopping pebbles outside of its orbit it cannot efficiently accrete pebbles anymore so its solid budget is confined. Moreover, the local change in surface density (the gap) modulates the strength of planet-disk interactions that drive a growing planet's disk-driven migration. Classically, migration regimes are divided into a type-I regime, when the planet's mass is low enough (typically up to a few ten's of an Earth's mass for a typical protoplanetary disk) that it does not modify the disk's underlying surface density much, and a type-II regime, where the planet starts carving a significant enough gap \citep{2006Icar..181..587C,2018ApJ...861..140K,2018A&A...617A..98R}, influenced also by the accretion of gas onto the growing planet (e.g.\ \citealt{2017Icar..285..145C,2020A&A...643A.133B}). 

One thus needs to understand the formation of gaps on quantitative grounds. We note that the gap depth is known to depend not only on the planet's mass but on disk properties as well, namely its aspect ratio and turbulent viscosity \citep{2006Icar..181..587C,2018ApJ...861..140K}. In particular, in the presence of a planet of a given mass, thinner and less viscous disks will respond with a deeper gap than thicker and more viscous disks. 
In fact, the processes that were thought to drive turbulent viscosities, such as the Magneto-Rotational Instability (MRI, \citealt{1991ApJ...376..214B}), are quenched in large portions of the disk midplanes where planets form, and the remaining hydro-instabilities generate viscosities that are at least an order of magnitude lower than expected (\citealt{2014ApJ...789...77L,2021ApJ...915..130P,2018ApJ...869..127B}, see \citealt{2023ASPC..534..465L} for a review). The analysis of observed disks also shows that results are best reproduced for similar low viscosities (see for example HL Tau and Oph163131 respectively by \citealt{2016ApJ...816...25P} and \citealt{2022ApJ...930...11V}).
Thus, even a low-mass planet that would traditionally be considered in the type-I regime may start opening a partial gap. For this reason, as convenient as the separation of migration regimes may be, we must be able to accurately describe modes of planetary-disk interactions that lie in between the two classical extremes. We note that planets that would fall in such transitional regimes have masses of a few to a few ten's of Earth's mass (so called Mini-Neptunes or Super-Earths, depending on whether they feature a thin gaseous atmosphere or not). Such planets do not exist in our own Solar System but appear to be the most common type of exoplanet in the galaxy \citep{2011arXiv1109.2497M,2013ApJ...766...81F,2013PNAS..11019273P,2015ARA&A..53..409W,
2018ApJ...860..101Z,2021AJ....161...16H,
2023ASPC..534..839L}, and represent the cores of giant planets in the core accretion model (\citealt{1996Icar..124...62P}).

Planet-disk interactions can be understood as a combination of changes in the planet's orbit's size, shape and inclination with respect to the disk mid-plane; these are usually referred to as a planet's migration (which is typically inward, thus associated to a damping of the planet's semi-major axis\footnote{
More precisely, of the norm of the angular momentum, see Sect.\ \ref{subsec:OEDT}.}), eccentricity damping and inclination damping, respectively. Concerning the migration efficiency, \cite{2018ApJ...861..140K} has shown that, for circular and coplanar orbits, the transition between migration speeds in the classical type-I and type-II regime is modulated linearly by the depth of the gap carved by the planet.
In a previous paper \citep{2023A&A...670A.148P}, we showed using high-resolution 2D locally isothermal hydrodynamical simulations that, for non-inclined planets, eccentricity damping efficiencies follow a similar trend, with a linear dependence on the gap depth (whose slope and intercept depend on the eccentricity). This fact is supported by theoretical grounds based on the gap profile opened by the planet and our understanding of the so-called eccentricity waves responsible for driving the eccentricity evolution of a planet embedded in a disk \citep{1980ApJ...241..425G,2004ApJ...602..388T,1988Icar...73..330W, 2008EAS....29..165M,2015ApJ...812...94D}. 
We found that $e$-damping efficiencies can be significantly lower than in the case of shallow gaps \citep{2004ApJ...602..388T,2008A&A...482..677C}; this finding bridges the gap between the classical regime of eccentricity damping that is typically associated to low-mass planets and the eccentricity pumping that is observed for high-mass planets in the type-II regime \citep{2001A&A...366..263P,2006A&A...447..369K,2013A&A...555A.124B,2015ApJ...812...94D}.
We also obtained a fitting formula to predict the gap depth of a planet of a given mass embedded in a disk with a given aspect ratio and viscosity that is formally similar to the one from \cite{2018ApJ...861..140K}, but gives a more accurate prediction for partial-gap opening planets in the low-viscosity regime, not probed by \cite{2018ApJ...861..140K}.
In this paper, we extend our analysis to inclination damping efficiencies by using high-resolution 3D locally isothermal hydrodynamical simulations, and allowing the planet to lie on inclined orbits with respect to the disk mid-plane. At the same time, we refine our results on the eccentricity damping since there are known differences in $e$-damping efficiencies between 2D and 3D \citep{2004ApJ...602..388T}. The fitting formula for the gap depth for partial-gap opening planets in low-viscosity disks is also investigated in the 3D case. Finally, we use the resulting gas surface density profiles obtained in our simulations to characterise the final mass that such planets would reach by accreting pebbles in their protoplanetary disk (the pebble isolation mass) and compare the outcome with known results from the literature \citep{2018A&A...612A..30B}.

The paper is organised as follows. Section \ref{sec:DiscModel} describes our disk model and the setups used in our hydro-dynamical simulations. Section \ref{sec:Methods} gives a comprehensive description of planet-disk interaction schemes in hydro- and $N$-body simulations: this includes how orbital damping timescales are extracted from hydro simulations and how they are implemented in $N$-body codes, including the transition from the type-I to the type-II migration regime. Section \ref{sec:Results} describes the main results of our hydro simulations, yielding simple fitting formulas for the (partial) gap opened by a planet, and for the eccentricity and inclination damping timescales when planets open partial gaps in their surrounding disks, as a function of the orbital eccentricity and inclination. In Section \ref{sec:Discussion} we compare our work to previous results on pebble isolation mass scaling laws, to yield a complete description of the dynamical interactions of partial-gap opening planets with their surrounding protoplanetary disks, and we discuss the implications of our orbital damping formulas in the context of population synthesis models. Finally, Section \ref{sec:Conclusion} summarises our results.

\section{Disc Model}\label{sec:DiscModel}
We consider a gaseous 3D disk extending from 0.5 to 2 AU around a $M_* = M_\odot$ star. We take a power-law profile for the surface density $\Sigma(r) = \Sigma_0 (r/r_0)^{-\alphaSigma}$, a constant aspect ratio $H/r=h=h_0$ (flaring index $\betah=0$), and a constant $\alphaturb$ turbulence viscosity parameter for the well-known alpha-prescription $\nuvisc=\alphaturb H^2 \Omega_\Kepl$ \citep{1973A&A....24..337S}. 
Assuming a constant accretion rate $\dot M_{\gas,\accr} = 3\pi\nu\Sigma_\gas = 3 \pi \alphaturb h ^ { 2 } r ^ { 2 } \Omega_\Kepl\Sigma_\gas$ sets $\alphaSigma = 0.5$.
Like in our previous paper \cite{2023A&A...670A.148P}, we assume that the locally isothermal approximation is valid, that is, we prescribe a temperature profile $T(r) = T_0 (r/r_0)^{-\betaT}$, where $\betaT = 1-2 \betah = 1$ (note that $e$- and $\I$-damping are similar in isothermal and fully-radiative discs, \citealt{2010A&A...523A..30B,2011A&A...530A..41B}). 
Disk winds are neglected.
We then add a planet at a distance of 1 AU on a fixed orbit (we show in Appendix \ref{apx:FreePlanets} that the results on orbital damping efficiencies obtained with a planet on a fixed orbit are comparable to the case of a planet let free to evolve in the disc). The mass of the planet increases smoothly from an initial value of 0 to its final masses (which varies across simulations, see below) over the course of 50 orbits. We consider planetary masses that are always below the thermal mass $m_\mathrm{th} = h^3 M_*$, so that the disk-planet tidal perturbation does not drive local nonlinear shocks and can be treated linearly \citep{1986ApJ...309..846L}.

We thus have surface density and temperature profiles fixed across all simulations, while $\alphaturb$, $h$ and $\mpl/M_*$ are left as free parameters. We consider $\alphaturb$ values between $3.16\times 10^{-5}$ and $10^{-3}$, since in the MRI-dead zone a residual turbulent viscosities can arise by purely hydrodynamical instabilities, with $\alphaturb$ of order $10^{-4}$ (e.g.\ \citealt{2021ApJ...915..130P, 2020ApJ...897..155F,2023ASPC..534..465L}), and observational constraints determine $\alphaturb$ in disks to range between $10^{-5}$ and $10^{-2}$ \citep{2016ApJ...816...25P,2017ApJ...837..163R, 2018ApJ...869L..46D, 2017ApJ...843..150F, 2018ApJ...856..117F,2022ApJ...930...11V}. We take the disk's aspect ratio $h \in \{0.04, 0.05, 0.06\}$ and planet masses in the Super-Earth/Mini-Neptune range, $\mpl/M_*\in \{1 \times 10^{-5}, 3\times 10^{-5},6\times 10^{-5}\}$.
Compared to the 2D case, for a given disk structure and planetary mass we need to vary not only $e/h$ but also $\I/h$ (where $e$ is the planet's eccentricity and $\I$ is the inclination); moreover, 3D simulations are more costly than 2D ones. For this reason, in order to spare computational resources, we do not investigate a full grid of parameters $\{\mpl/M_*,\alphaturb,h\}$ like we did in \cite{2023A&A...670A.148P}, but we only consider setups which allow us to reach different gap depths at relatively equally spaced intervals to obtain a 3D version of the results obtained in \cite{2023A&A...670A.148P}. 

We give in Table \ref{tbl:parameters} the list of all our disk setups. 
For each of the setups in the top entries, we varied the values of $e/h$ and $\I/h$ independently in $\{0, 0.25, 0.5, 1\}$.
We also run simulations for additional setups in the non-eccentric and non-inclined cases (bottom entries of Table \ref{tbl:parameters}), which we used to update our prediction for the gap depth. These total to 167 high-resolution 3D simulations.
We do not consider higher values for the eccentricity and inclination for similar reasons as our 2D paper: the analytical formulas that we wish to compare our results to start breaking down at higher $e/h$ and $\I/h$; small single planets in such mass range have their eccentricities/inclinations damped by the disk so they are not expected alone to reach large $e$ or $\I$ \citep{2010A&A...523A..30B,2011A&A...530A..41B,2008A&A...482..677C}; when multiple planets interact in a disk, e.g.\ by capturing in resonance via convergent migration, the expected capture eccentricities are of order $h$ \citep{2005MNRAS.363..153P,2008A&A...483..325C,2014AJ....147...32G,2015ApJ...810..119D,2018CeMDA.130...54P}, and the inclinations can be excited only by second order effects. 
Along the simulation, the planet's orbit (semi-major axis $a$, eccentricity $e$ and inclination $\I$) is kept fixed, as well as its mass $m_\pln$ (except for the initial mass taper).

\begin{center}
\begin{table}
\centering
\begin{tabular}{ c l } 
 \hline
 Parameters & Values  \\ \hline \hline 
 $\{\mpl/M_*,\alphaturb, h\}$ 
    & $\{1\times 10^{-5}, 1\times 10^{-3}, 0.04\}$\\
    & $\{1\times 10^{-5}, 1\times 10^{-4}, 0.04\}$\\
    & $\{1\times 10^{-5}, 1\times 10^{-4}, 0.05\}$\\
    & $\{1\times 10^{-5}, 1\times 10^{-4}, 0.06\}$\\
    & $\{3\times 10^{-5}, 1\times 10^{-3}, 0.04\}$\\
    & $\{3\times 10^{-5}, 1\times 10^{-3}, 0.05\}$\\
    & $\{6\times 10^{-5}, 1\times 10^{-3}, 0.04\}$\\
    & $\{6\times 10^{-5}, 1\times 10^{-3}, 0.05\}$\\
    & $\{6\times 10^{-5}, 3.16\times 10^{-4}, 0.04\}$\\
    & $\{6\times 10^{-5}, 3.16\times 10^{-5}, 0.05\}^{(*)}$ 
 \\\hline 
 $e/h$      & $\{0, 0.25, 0.5, 1\}$ for all setups above\\
 $\I/h$      & $\{0, 0.25, 0.5, 1\}$ for all setups above\\ \hline \hline
 $\{\mpl/M_*,\alphaturb, h\}$ 
    & $\{1\times 10^{-5}, 3.16\times 10^{-4}, 0.06\}$\\  
    & $\{3\times 10^{-5}, 1\times 10^{-4}, 0.04\}$\\  
    & $\{3\times 10^{-5}, 1\times 10^{-4}, 0.05\}$\\   
 (only for gap depth,    
    & $\{3\times 10^{-5}, 3.16\times 10^{-4}, 0.04\}$\\
  $e/h=\I/h=0$)  
    & $\{3\times 10^{-5}, 3.16\times 10^{-4}, 0.06\}$ \\ 
    & $\{6\times 10^{-5}, 3.16\times 10^{-4}, 0.05\}$\\ 
    & $\{6\times 10^{-5}, 3.16\times 10^{-4}, 0.06\}$\\ 
    \hline 
\end{tabular}
\caption{Set of parameters of our simulations. In the case marked with an asterisk, the disk was unstable.}
\label{tbl:parameters}
\end{table}
\end{center}

We note that in our simulations the reference frame is centered on the star and indirect forces should be considered. Similarly to our 2D study, we apply indirect forces to all the elements that feel a direct gravitational force: the planet feels the indirect force due to its own gravity as well as that of the disk; the disk feels indirect forces from the planet; the indirect forces of the disk onto itself is not included since we do not consider the disk's self-gravity.

Like in our previous paper \cite{2023A&A...670A.148P}, we run our numerical experiments using the \texttt{fargOCA} code (fargo with {\bf C}olatitude {\bf A}dded; \citealt{2014MNRAS.440..683L})\footnote{The code can be found at: \url{https://gitlab.oca.eu/DISC/fargOCA}}, which is a 3D extension of the \texttt{fargo} code \citep{2000A&AS..141..165M}, parallelised using a hybrid combination of MPI and Kokkos \citep{CARTEREDWARDS20143202,9485033}.
Code units are $G=M_*=1$, and the unit of distance $r_0=1$ is arbitrary when expressed in AU. We used 512 grid cells with arithmetic spacing in radius for a disk extending from 0.5 to 2 AU, 2000 cells in azimuth for the full $(0,2\pi)$, and 70 cells in zenith for a disk with colatitude of $83^\circ$. 
This gives square-ish cells at the location of the planet with $\delta r\simeq \delta\phi \simeq \delta\theta \simeq 0.003$.
This resolution is similar to our 2D runs, which we achieved by considering a slightly narrower radial domain in order to maintain a manageable total number of grid cells. Our disk is however still larger compared to the ones used in \cite{2017MNRAS.471.4917J}, while at the same time achieving a higher resolution. 
This ensures that, even for the smallest planetary masses, we are resolving six cells in a half horseshoe width of such planets, which is needed in order to properly resolve the co-rotation torque \citep{2011MNRAS.410..293P, 2014MNRAS.440..683L}. 
We performed resolution convergence tests as described in Appendix \ref{apx:Convergence}.
We used a smoothing length for the potential of the planet of $r_\mathrm{sm} = s_\mathrm{sm} R_\mathrm{H}$ with $s_\mathrm{sm} = 0.5$. Finally, we used radially evanescent and vertically reflecting boundary conditions \citep{2006MNRAS.370..529D}.

\section{Methods}\label{sec:Methods}

\subsection{Orbital elements damping timescales}\label{subsec:OEDT}
At regular time intervals, the \texttt{fargOCA} code outputs the (direct) gravitational force felt by the planet from the disk and the force felt by the star from the disk (indirect term). This indirect force emerges because of the asymmetry in the gas density resulting from the gas' response to the presence of the planet. Since our simulations are performed in a non-inertial astrocentric frame of reference, this force will result in an indirect (fictitious) response force felt by the planet, so we need to add it to the direct force felt by the planet from the gas. The resulting force $\bfF_{\disk\to\pln}$ describes the sum of direct and indirect planet-disk interactions, and thus the true force felt by the planet in an inertial reference frame. As in \cite{2023A&A...670A.148P}, we use orbit-averaged forces, where the average is done over 20 points along the planet's orbit.

Following \cite{1976AmJPh..44..944B}, we decompose the force $\bfF_{\disk\to\pln}$ into three components: 
\begin{equation}\label{eq:PerturbingForceDecomp}
    \bfF_{\disk\to\pln} = R \bfe_R + T \bfe_T + N \bfe_N,
\end{equation}
where $\bfe_R$, $\bfe_T$ and $\bfe_N$ represent an orthonormal vector triad such that $\bfe_R$ is in the direction of $\bfr_\pln$, $\bfe_T$ is inside the orbital plane and transverse to $\bfr_\pln$, and $\bfe_N$ is perpendicular to the orbital plane in the direction $\bfe_R\times\bfe_T$, which is also the direction of the (orbital) angular momentum vector $\bfANGMOM = \redmass \bfr_\pln \times \bfv_\pln$.
Here, $\bfr_\pln$ is the position of the planet, $\bfv_\pln:=\dot\bfr_\pln$ is its velocity, and $\redmass = (m_\pln M_*)/(M_*+m_\pln) \approx m_\pln$ is the reduced mass of the planet. We also introduce $\ANGMOM:=\|\bfANGMOM\|$ the norm of the angular momentum vector, given by
\begin{equation}
    \ANGMOM = \redmass \sqrt{\GM a (1-e^2)},
\end{equation}
where $\GM = \GravC (M_*+m_\pln) \approx \GravC M_*$ is the reduced gravitational parameter, and $a$ and $e$ are the semi-major axis and eccentricity of the planet's orbit. The norm of the angular momentum is independent of the orbit's inclination $\I$, which only dictates the orientation of $\bfANGMOM$ (see below). Finally, we introduce the (orbital) energy
\begin{equation}
    E := -\frac{\GM \redmass}{2a}.
\end{equation}

We need to determine how the different components of the perturbing force \eqref{eq:PerturbingForceDecomp} translate into orbital elements damping. This damping is typically defined through damping timescales $\tau_a$, $\tau_e$ and $\tau_\I$ via
\begin{align}
\label{eq:tauaDEF}
    \frac{\dot{a}}{a}   &=: -\frac{1}{\tau_a},\\
\label{eq:taueDEF}
    \frac{\dot{e}}{e}   &=: -\frac{1}{\tau_e},\\
\label{eq:tauincDEF}
    \frac{\dot{\I}}{\I}   &=: -\frac{1}{\tau_\I};
\end{align}
like in the 2D case we also define the migration timescale $\tau_\mig$ via
\begin{equation}\label{eq:taumigDEF}
    \frac{\dot{\ANGMOM}}{\ANGMOM} =: -\frac{1}{\tau_\mig},
\end{equation}
that is, the damping timescale of the (norm of the) angular momentum, which is different from the semi-major axis evolution timescale. 
Only forces inside the orbital plane ($R \bfe_R + T \bfe_T$) can change the orbit's shape (semi-major axis and eccentricity, and thus the norm of the angular momentum), while forces perpendicular to it will change its orientation, i.e.\ the orbit's inclination. 

Inside the orbital plane, the evolution of $a$ and $e$ due to disk-planet perturbative forces is similar to the 2D case in \citealt{2023A&A...670A.148P}.
The time derivatives of $\bfANGMOM=\ANGMOM \bfe_N$ and of its norm $\ANGMOM$ are given by the torque:
\begin{align}
\label{eq:TorqueVectorialForm}
    \dot\bfANGMOM    &= \bfr_\pln \times \bfF_{\disk\to\pln} = r_\pln(T \bfe_N - N \bfe_T),\\
\label{eq:TorqueNorm}
    \dot\ANGMOM      &=\dot\bfANGMOM\cdotp\bfANGMOM/\ANGMOM = r_\pln T=:\Gamma,
\end{align}
where $r_\pln = \|\bfr_\pln\|$ and only $r_\pln T\bfe_N$ contributes to the change in $\ANGMOM$ since $-r_\pln N \bfe_T$ is perpendicular to $\bfANGMOM$ and only changes its direction (see below). 
We then consider the power
\begin{equation}
    P := \bfF_{\disk\to\pln} \cdotp \bfv_\pln,
\end{equation}
which represents a change in orbital energy, $P=\dot E$. Using these expressions, one calculates (see e.g.\ \citealt{2023A&A...670A.148P} for an explicit derivation)
\begin{align}
\label{eq:taumig.from.torque}
    \tau_\mig   &= -\frac{\ANGMOM}{\Gamma},\\
    \tau_a      &= \frac{E}{P},\\
\label{eq:taue.wrt.taumig-taua}
    \tau_e      &= C(e) \left(-\frac{1}{\tau_\mig} + \frac{1}{2\tau_a}\right)^{-1},
\end{align}
where 
\begin{equation}
C(e) = \frac{e^2}{1-e^2},
\end{equation}
and $C(e)\approx e^2$ for small eccentricities.
For a circular orbit $\tau_\mig = 2\tau_a$.

To obtain $\tau_\I$, we follow \cite{1976AmJPh..44..944B} equation (32) (see also \citealt{2011A&A...530A..41B}) and write
\begin{equation}
    \frac{\D\I}{\D t} = \frac{r_\pln N \cos\theta_\pln}{\ANGMOM},
\end{equation}
where $\theta_\pln$ is the planet's true longitude, and average this quantity over one orbit. Only $N$ appears here because forces in the orbital plane cannot change the plane's orientation. This yields
\begin{align}
    \tau_\I   &= -\frac{\I}{\left\langle \frac{\D\I}{\D t} \right\rangle}.
\end{align}

\subsection{3D planet-disk interactions in N-body codes}
$N$-body codes implement type-I migration using timescales $\tau_\mig$, $\tau_e$ and $\tau_\I$ to define accelerations onto the planet given by \citep{2000MNRAS.315..823P}
\begin{align}
    \label{eq:tau_mig.PL2000}
    \bfa_\mig   &= -\frac{1}{\tau_\mig}\bfv_\pln,\\
    \label{eq:tau_e.PL2000}
    \bfa_e      &= -2\frac{(\bfv_\pln\cdotp\bfr_\pln)}{r_\pln^2 \tau_e}\bfr_\pln,\\
    \label{eq:tau_inc.PL2000}
    \bfa_\I      &= -2\frac{(\bfv_\pln\cdotp\bfk)}{\tau_\I}\bfk,
\end{align}
where $\bfr_\pln$ and $\bfv_\pln$ are the planet's position and velocity, and $\bfk$ is the unit vector in the vertical direction. These accelerations directly damp canonical momenta (the three Delaunay action variables, e.g.\ \citealt{2002mcma.book.....M}) associated to the orbital elements $a$, $e$ and $\I$ as we show below.\\

The first equation describes a torque, i.e.\ a change in the angular momentum vector as
$\dot\bfANGMOM
= \redmass \bfr_\pln \times \ddot{\mathbf{r}}_\pln$ = $-\redmass (\bfr_\pln \times \bfv_\pln)/{\tau_\mig} \equiv - \bfANGMOM/\tau_\mig$. This implies that its norm $\ANGMOM$ (the first Delaunay action) evolves according to \eqref{eq:taumigDEF}. Since $\bfa_\mig\parallel\bfv_\pln$, the resulting force lies on the orbital plane. 

Equation \eqref{eq:tau_e.PL2000} also represents a force that lies on the orbital plane, but it has zero torque since $\bfa_e\parallel\bfr_\pln$, so it does not contribute to  a change in $\ANGMOM$. 
By applying \eqref{eq:tau_e.PL2000} over an orbit, the quantity $E=(1-e^2)^{-1/2}-1$ is damped exponentially over a timescale $\tau_e/2$ \citep{2023A&A...670A.148P}. $E$ is the ratio between the Angular Momentum Deficit (AMD) $\GAMMONA = \redmass \sqrt{\GM a}(1-\sqrt{1-e^2})$ of the planet \citep{1997A&A...317L..75L} and the norm of the angular momentum vector $\ANGMOM$, and it represents the second Delaunay action. Note that $E\simeq e^2/2$ for small $e$'s, so that $\dot{E}/E = -2/\tau_e$ translates into $\dot{e}/e = -1/\tau_e$ for small $e$'s.
Therefore, at small $e$'s, Equation \eqref{eq:tau_e.PL2000} implements an exponential damping of the eccentricity over a timescale $\tau_e$ as described by equation \eqref{eq:taueDEF}.
The semi-major axis evolution described by \eqref{eq:tauaDEF} results from a combination of torque and $e$-damping with a timescale given by 
\begin{equation}\label{eq:taua.wrt.taumig-taue}
\tau_a = \left(\frac{1}{\tau_\mig/2} +  \frac{C(e)}{\tau_e/2}\right)^{-1},
\end{equation}
which is the equivalent of \eqref{eq:taue.wrt.taumig-taua}.

Equation \eqref{eq:tau_inc.PL2000} implements a force such that $\dot \ANGMOM_x/\ANGMOM_x = \dot \ANGMOM_y/\ANGMOM_y = -1/\tau_\I$ and $\dot \ANGMOM_z/\ANGMOM_z=0$, where $\ANGMOM_{\{x,y,z\}}$ are the component of $\bfANGMOM$. Using $\cos\I = \ANGMOM_z/\ANGMOM$ we introduce $I=1-\cos\I=1-\ANGMOM_z/\ANGMOM$. This quantity (up to a sign) represents the third Delaunay action. From the expressions of $\dot{\ANGMOM}_{\{x,y,z\}}$, one easily calculates that $I$ follows the evolution $\dot{I}/I=(-1/\tau_\I)(1-I)(2-I)$, which has the closed form solution $I(t)=1-(1+C\exp{\big[-2t/\tau_\I\big]})^{-1/2}$ where $C=\big(1/(1-I(0))^2-1)$. For small $I$'s, $\dot{I}/I\approx (-2/\tau_\I)$ so that $I$ gets exponentially damped over a timescale $\tau_\I/2$.
Moreover, $I \simeq \I^2/2$ for small $\I$'s (i.e.\ for small $I$'s), so $\dot{I}/I\approx-2/\tau_\I$ translates to $\dot\I/\I \approx -1/\tau_\I$ for small $\I$'s. Therefore, at small $\I$'s, Equation \eqref{eq:tau_inc.PL2000} implements an exponential damping of the inclination over a timescale $\tau_\I$ as described by equation \eqref{eq:tauincDEF}.\\

For inclined orbits, the acceleration $\bfa_\I$ in Equation \eqref{eq:tau_inc.PL2000} has components both inside and perpendicular to the orbital plane. For this reason, it also modifies the norm of the angular momentum, and therefore includes an additional unwanted (albeit small) change in the orbit's shape ($\dot\ANGMOM\neq0$).
To overcome this nuisance, one can define a modified perturbing acceleration given by
\begin{align}
    \label{eq:tau_inc.PL2000.New}
    \tilde{\bfa}_\I	&= -2\frac{(\bfv_\pln \cdotp \bfk)}{\tau_\I}\bfe_N,
\end{align}
where $\bfe_N = \left(\sin\I\sin\OMEGONA, -\sin\I\cos\OMEGONA,\cos\I\right)^\intercal$ is again the versor orthogonal to the orbital plane and parallel to $\bfANGMOM$. With this modified acceleration, $\dot \ANGMOM_x/\ANGMOM_x = \dot \ANGMOM_y/\ANGMOM_y = -\ANGMOM_z/(\ANGMOM \tau_\I)$ (they gain a factor of $\cos\I=\ANGMOM_z/\ANGMOM$), while $\dot\ANGMOM = 0$ as desired. From this, one easily calculates that, under \eqref{eq:tau_inc.PL2000.New}, the quantity $I=1-\cos\I$ follows the evolution $\dot{I}/I=(-1/\tau_\I)(2-I)$, which has the closed form solution $I(t)=2(1+C \exp{\big[2t/\tau_\I\big]})^{-1}$ where $C=(2-I(0))/I(0)$. For small $I$'s, $\dot{I}/I\approx (-2/\tau_\I)$ so that also in this case $I$ gets exponentially damped over a timescale $\tau_\I/2$. Thus, like before, Equation \eqref{eq:tau_inc.PL2000.New} implements at small $\I$'s an exponential damping of the inclination over a timescale $\tau_\I$ as described by equation \eqref{eq:tauincDEF}, but without introducing a spurious damping of $\ANGMOM$. This $N$-body implementation is thus better suited to be in line with the output of hydrodynamical simulations in the case of inclined orbits (at the expense of calculating the versor $\bfe_N = \bfANGMOM/\ANGMOM$).

\subsubsection{Type-I forces and transition to the type-II regime}\label{subsec:TIForces}
Analytical formulas for the damping timescales $\tau_\mig$, $\tau_e$ and $\tau_\I$ for various migration regimes have been the subject of a number of works and we briefly summarise them here in the context of this work.

Much attention has been given to the expression of the torque $\Gamma$ (i.e.\ of $\tau_\mig$ by Eq.\ \eqref{eq:taumig.from.torque}) in the circular and non-inclined case for low-mass, non-gap opening planets (e.g.,\ \citealt{2002ApJ...565.1257T,2008A&A...482..677C,2011MNRAS.410..293P,2017MNRAS.471.4917J}). The total torque is typically split into a Lindblad component $\Gamma_\Lind$ (usually negative) and a corotation component $\Gamma_\Corot$ (usually positive, but prone to saturation). The total torque $\Gamma_{\tot,\mathrm{I}} = \Gamma_\Lind + \Gamma_\Corot$ provides a nominal type-I migration timescale $\tau_{\mig,\mathrm{I}}$, which depends explicitly on the disk structure, particularly on the surface density profile ($\alphaSigma$), temperature profile ($\betaT$), aspect ratio $h$ and viscosity $\nuvisc$ at the location of the planet, as well as the planet's mass. 
The transition from non-gap opening (classic type-I) to partial gap opening planets to type-II planets has been investigated by \cite{2018ApJ...861..140K}, who found that 
the type-I migration timescales is modulated linearly by the depth of the gap carved by the planet in the disk:
\begin{equation}\label{eq:type-ItoIIMigTransition}
    \tau_{\mig,\mathrm{II}} = \left(\frac{\Sigma_{\min}}{\Sigma_0}\right)^{-1} \tau_{\mig,\mathrm{I}}.
\end{equation}
Here $\Sigma_{\min}/\Sigma_0$ measures the gap depth, where $\Sigma_{\min}$ is the minumum of the (azimuthally averaged) surface density $\Sigma(r)$ near the location of the planet $r_\pln$, while $\Sigma_0$ is the unperturbed disk surface density. \cite{2018ApJ...861..140K} also provided a fitting formula for $\Sigma_{\min}/\Sigma_0$:
\begin{equation}\label{eq:KanagawaGapDepth}
    \Sigma_{\min}/\Sigma_0\simeq \frac{1}{1+0.04 K},
\end{equation}
where
\begin{equation}\label{eq:KanagawaK}
    K: = q^2 h^{-5} \alphaturb^{-1}
\end{equation}
is a dimensionless parameter.
After the gap is considered to have been fully opened \citep{2006Icar..181..587C,2018ApJ...861..140K}, migration occurs in the type-II migration regime \citep{1986ApJ...309..846L, 2018A&A...617A..98R} and is beyond the scope of this work (we provide a measure of this limit in subsect.\ \ref{subsec:GPEV}).
For eccentric and inclined orbits, the approach taken by many authors is to modulate the Lindblad and corotation torques by factors $\Delta_\Lind$ and $\Delta_\Corot$ that depend on $e$ and $\I$ \citep{2013A&A...553L...2C, 2013A&A...558A.105P, 2014MNRAS.437...96F}, to obtain a torque $\Gamma_{\tot,\mathrm{I}} = \Delta_\Lind \Gamma_\Lind + \Delta_\Corot \Gamma_\Corot$ for eccentric and inclined planets that can be used in population synthesis models \citep{2017MNRAS.470.1750I, 2021A&A...650A.152I, 2021A&A...656A..69E}.

The disk-driven eccentricity and inclination evolution has also been the subject of many works (e.g.\ \citealt{1983Icar...53..185S,1988Icar...73..330W,1994ApJ...423..581A,1994Icar..110...95W,2004ApJ...602..388T,2008A&A...482..677C,2011A&A...530A..41B,2023A&A...670A.148P}). In particular, \cite{2004ApJ...602..388T} gave the first expression for $\tau_e$ and $\tau_\I$, which was extended by \cite{2008A&A...482..677C} to non-vanishing eccentricities and inclinations, by fitting the orbital evolution of a planet embedded in a 3D disk. Their fits yield
\begin{align}
\label{eq:taue_CN2008}
    \tau_{e, \mathrm{CN2008}} &= \frac{\tau_\mathrm{wave}}{0.780}\left[1-0.14\left(\frac{e}{h}\right)^2+0.06\left(\frac{e}{h}\right)^3
    +0.18 \left(\frac{e}{h}\right)\left(\frac{\I}{h}\right)^2
    \right],\\
\label{eq:tauinc_CN2008}
    \tau_{\I, \mathrm{CN2008}} &= \frac{\tau_\mathrm{wave}}{0.544}\left[1-0.30\left(\frac{\I}{h}\right)^2+0.24\left(\frac{\I}{h}\right)^3
    +0.14 \left(\frac{e}{h}\right)^2\left(\frac{\I}{h}\right)
    \right].
\end{align}
Here $\tau_\mathrm{wave}$ is the typical type-I damping timescale \citep{2004ApJ...602..388T},
\begin{equation}
    \tau_\mathrm{wave} = \left(\frac{M_{\odot}}{m_\mathrm{pl}}\right) \left(\frac{M_\odot}{\Sigma_\mathrm{gas,pl} a_\mathrm{pl}^2}\right) h_\pln^4 \Omega_{\Kepl,\mathrm{pl}}^{-1},\label{eq:tauwave}
\end{equation}
were $\Omega_\Kepl$ is the Keplerian orbital frequency and quantities with a subscript pl are evaluated at the position of the planet.
These formulas are the most commonly used $e$- and $\I$-damping prescription in the population synthesis literature (e.g.\ \citealt{2017MNRAS.470.1750I, 2021A&A...650A.152I, 2021A&A...656A..69E}, which use a mix of \cite{2011MNRAS.410..293P}'s formula for the torque plus \cite{2008A&A...482..677C}'s formulas \eqref{eq:taue_CN2008} and \eqref{eq:tauinc_CN2008}). However, they are only valid for non-gap opening planets ($\Sigma_{\min}/\Sigma_0\simeq1$). In our preliminary 2D investigation \citep{2023A&A...670A.148P}, we extended the formula for $\tau_e$ to partial gap opening, eccentric but non-inclined planets, and found that, like $\tau_\mig$, also $\tau_e$ is modulated by a linear function of $\Sigma_{\min}/\Sigma_0$, whose slope and intercept depend on $e/h$. In this work, we further extend this study to 3D simulations where planets are allowed to reside on inclined orbits too.

\section{Results}\label{sec:Results}
\begin{figure}[t!]
\centering
\includegraphics[width= 0.45\textwidth]{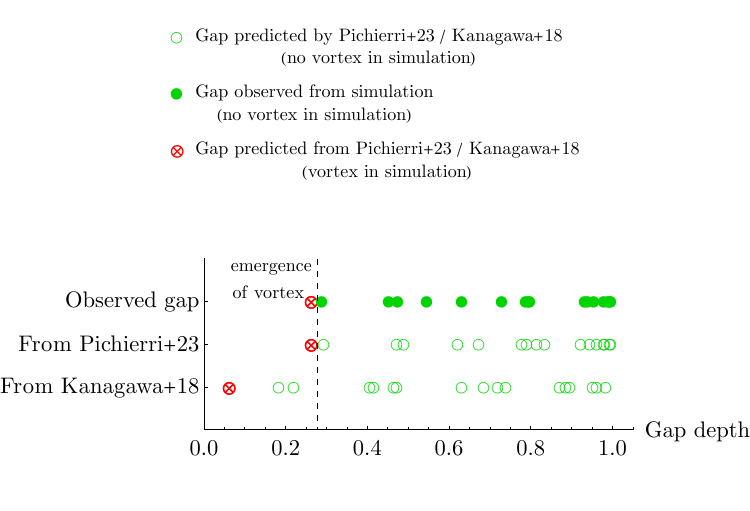}
    \caption{
    Gap depth measured for the set of simulations of Table \ref{tbl:parameters}. When no vortex was observed, we mark both the observed gap depth and the predicted one with a green circle (filled and unfilled respectively). Predicted values are obtained with the prescription from \cite{2018ApJ...861..140K} and from our 2D prescription from \cite{2023A&A...670A.148P}, which, in the low viscosity case not probed by \cite{2018ApJ...861..140K}, gives a better prediction of the gap depth observed also in 3D simulations. In one simulation (for the setup marked with an asterisk in Table \ref{tbl:parameters}) we observe a vortex. This case is marked with a red circle with a cross. In this case we can not measure a gap depth (because the gap continuously changes); therefore, in the observed gap line, we simply use the value predicted in \cite{2023A&A...670A.148P}. A dashed vertical line marks the approximate location where the transition between no vortex and vortex lies, which is similar to the one observed in 2D simulations \citep{2023A&A...670A.148P}.
    }
\label{fig:VortexLowerLimitPlot}%
\end{figure}
%

   \begin{figure*}
   \centering
   \includegraphics[height = 0.35\textwidth]{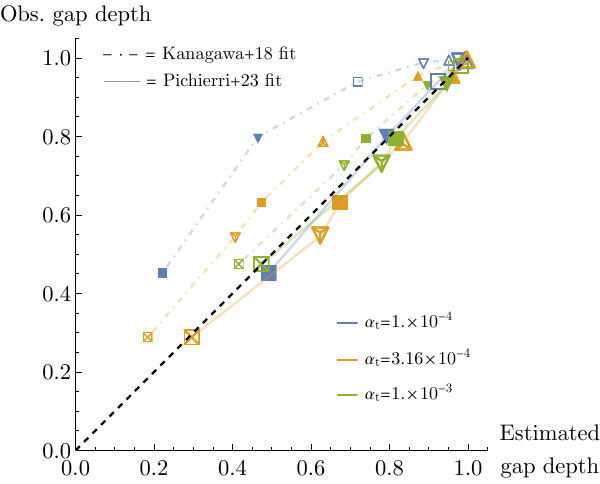} 
   \includegraphics[height = 0.35\textwidth]{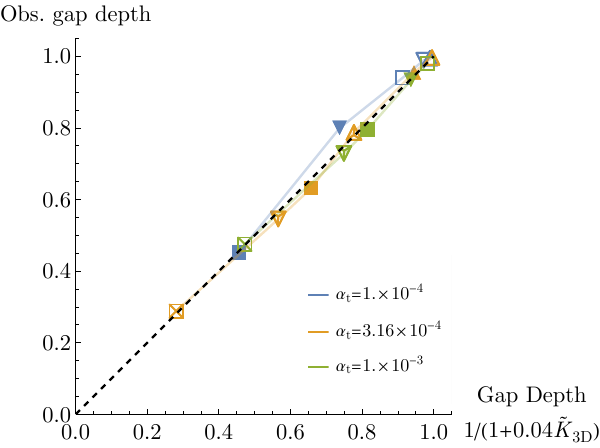} 
   \caption{
   Panel (a): Gap depth prediction (either from \citealt{2018ApJ...861..140K} or from \citealt{2023A&A...670A.148P}, indicated by small markers linked together by dot-dashed lines or larger markers linked by continuous lines, respectively) versus the observed 3D gap depth. Panel (b): New 3D gap depth prediction (Eqs.\ \eqref{eq:TildeKanagawaGapDepth3D}, \eqref{eq:tildeK3D}) versus observed 3D gap depth in the limit of circular and non-inclined orbits. In all panels different colours represent different levels of viscosity according to the legend, and symbols of different shapes are used to represent different aspect ratios and planetary masses: squares, downward-pointing triangles and upward-pointing triangles represent aspect ratios of 0.04, 0.05 and 0.06 respectively; empty, filled and crossed symbols represent $\mpl/M_*=10^{-5}$, $3\times 10^{-5}$ and $6\times 10^{-5}$ respectively.
   }
    \label{fig:PredictedVsObs_GapDepth}%
    \end{figure*}

\subsection{Gap opening and emergence of vortices}\label{subsec:GPEV}
Like in our 2D experiments, we run all $e=0$, $\I=0$ simulations up to at least 3000 planetary orbits (the integration time used in \citealt{2018ApJ...861..140K}) and recorded the final gap depth $\Sigma_{\min}/\Sigma_0$ carved by the planet.\footnote{
This is obtained by considering the surface density contrast $\Sigma_{t_\mathrm{max}}(r)/\Sigma_0(r)$ (where $\Sigma_{t_\mathrm{max}}$ is the azimuthally and vertically averaged surface density of the disk as a function of the radial distance $r$, at time $t_\mathrm{max}$, the total integration time), from which we mark the minimum value $\Sigma_{\min}/\Sigma_0 = \min_{r}\left[\left(\Sigma_{t_\mathrm{max}}/\Sigma_0\right)(r)\right]$, or take an average of the two minima found slightly exterior and interior to the planet's orbit.}
We note that the gap will clear over a timescale 5 to 10 times $(2/3)T_\mathrm{pl}(a_\mathrm{pl}/x_s)$, where $T_\mathrm{pl}$ is the orbital period and $x_s$ is the half-width of the horseshoe region \citep{2008EAS....29..165M}, which is at most $\sim 2000$ orbits in the cases we consider. 
The overdense regions located on each side of the gap will then spread radially to achieve a steady state, which is expected to happen over a longer (viscous) timescale. This timescale is too long to cover entirely in the context of hydro-dynamical simulations, so we adopt a practical approach where we prolong the integrations until the surface density does not vary significantly over time. 
Thus, in the lowest viscosity / thinnest disk runs, we extend the zero eccentricity and inclination run by an additional 1000 orbits. This ensures that, at the end of the simulations, the surface density always changes by less than 0.1\% over 50 orbits, meaning they have reached a quasi-steady state.

In \cite{2023A&A...670A.148P}, we found that an observed $\Sigma_{\min}/\Sigma_0\simeq0.25$ marked the transition from stable runs ($\Sigma_{\min}/\Sigma_0\gtrsim0.25$) and the appearance of vortices ($\Sigma_{\min}/\Sigma_0\lesssim0.25$). We confirmed that a similar behaviour is observed in 3D simulations. Indeed, we run a setup with $\mpl/M_*=6\times 10^{-5}$, $\alphaturb=3.16\times 10^{-5}$ and $h=0.05$, which has a predicted gap depth of $\Sigma_{\min}/\Sigma_0 \simeq 0.26$, which resulted in an unstable disk (marked with an asterisk in Table \ref{tbl:parameters} -- we note that for such a setup, all runs with different eccentricities and inclinations became unstable); instead, when $\mpl/M_*=6\times 10^{-5}$, $\alphaturb=3.16\times 10^{-4}$ and $h=0.04$, which has a predicted gap depth of $\Sigma_{\min}/\Sigma_0 \simeq 0.3$ (which is also the observed gap depth), the disk remained stable. We note that, like in the 2D case, the vortex appears for the lowest viscosity and the most massive planet, when the gap depth approaches $\sim 0.25$. Therefore the transition from type I migration to type II is also a transition from the no-vortex case to the vortex one but only because we are in a very low viscosity context.
Figure \ref{fig:VortexLowerLimitPlot} shows a diagram where we label the outcome of each simulation green when a vortex has not appeared and red when it has appeared. When a vortex has not appeared, we report both the observed gap depth (filled circle) and the predicted gap depth (unfilled circle) {from \cite{2018ApJ...861..140K} or \cite{2023A&A...670A.148P}}; when a vortex has appeared, we cannot use the simulations to observe a gap depth, and thus we only report the predicted gap depth from \cite{2023A&A...670A.148P}.
Our 2D-fitting formula from \cite{2023A&A...670A.148P} is similar to \cite{2018ApJ...861..140K}'s formula \eqref{eq:KanagawaGapDepth} (which was obtained for $\alphaturb=10^{-3}$), but provides a better fit to the data for lower viscosities, using a modified $K$ factor given by $\tilde{K}_\mathrm{2D} = 3.93 q^{2.3} h^{-6.14} \alphaturb^{-0.66}$.
We show in Figure \ref{fig:PredictedVsObs_GapDepth} (panel a) that this formula gives a good approximation for the gap depths obtained from 3D simulations as well. 
An improved fit that better matches the gap opening in 3D simulations is given by
\begin{equation}\label{eq:TildeKanagawaGapDepth3D}
    \Sigma_{\min}/\Sigma_0\simeq \frac{1}{1+0.04 \tilde{K}_\mathrm{3D}},
\end{equation}
where 
\begin{equation}\label{eq:tildeK3D}
    \tilde K_\mathrm{3D} = 28 q^{2.3} h^{-5.4} \alphaturb^{-0.72}.
\end{equation}
Figure \ref{fig:PredictedVsObs_GapDepth} panel (b) shows how this formula fits the data from 3D simulations.

When deeper gaps are carved by the planet and vortices appear, we cannot draw any definitive conclusion on the value of $\Sigma_{\min}/\Sigma_0$ or of the orbital damping timescales. However, like in \cite{2023A&A...670A.148P}, we note that planets that carve such deeps gaps are already considered in the type-II regime \citep{2018ApJ...861..140K,2006Icar..181..587C}, which is outside the scope of this work.
We also stress that we use the gap depth in the circular and non-inclined case as representative of the gap depth carved by planets on eccentric and/or inclined orbits. This is justified in the limit of our analysis as \cite{2007MNRAS.378..966H} showed that the gap carved by an eccentric planet is almost identical to the one carved by a planet on a circular orbit if $e<(\mpl/(3M_*))^{1/3}$, which is always the case for the setups considered here (see also \citealt{2010A&A...523A..30B}). More recently, \cite{2023MNRAS.518..439S} showed that the gap is fairly independent of the eccentricity if $e\lesssim h$. Finally, \cite{2011A&A...530A..41B} find little dependence of the gap depth for inclined orbits.

   \begin{figure*}
   \centering
   \includegraphics[width = 0.9\textwidth]{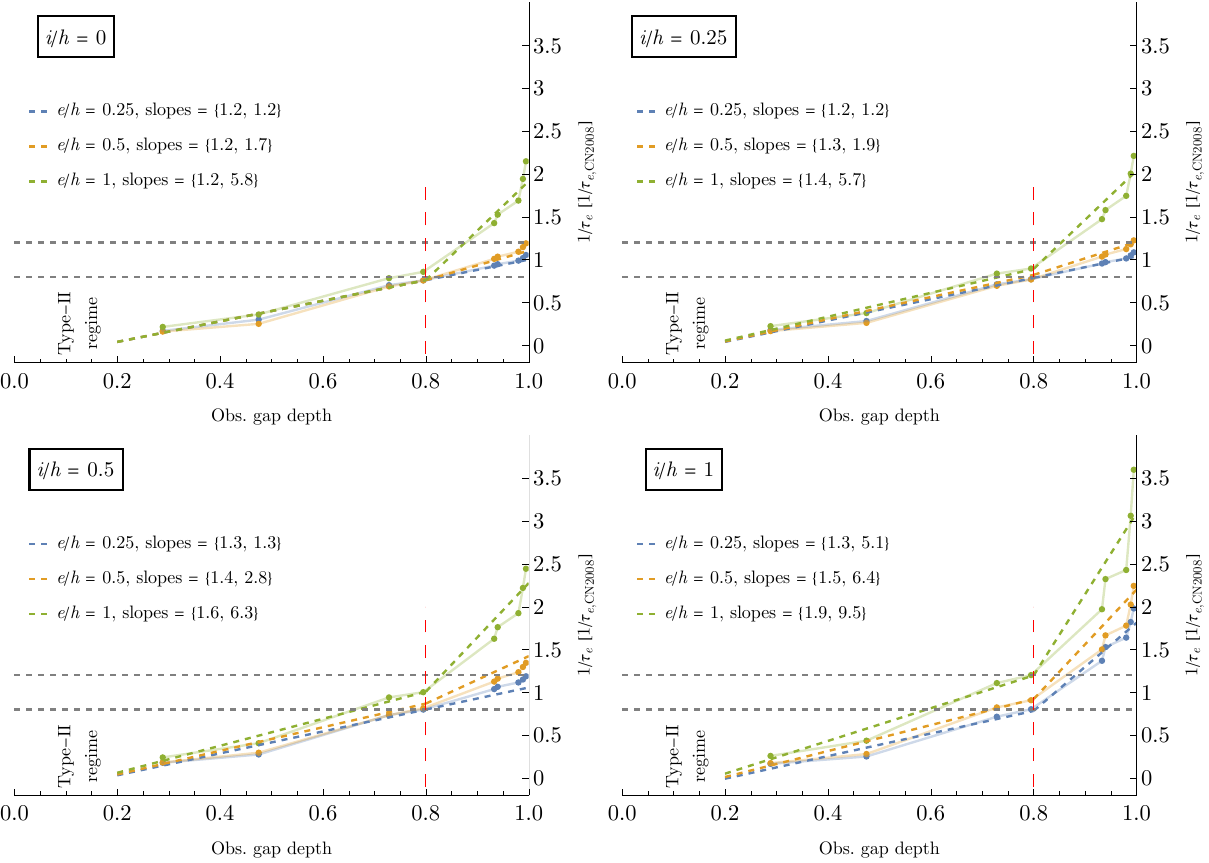} 
   \caption{
    Observed eccentricity damping efficiency versus observed gap-depth for all setups where no vortex emerged. In all panels, the $e$-damping efficiencies on the vertical axis are normalised by the expected value from \cite{2008A&A...482..677C} and are shown for different eccentricities $e/h \in \{0.25, 0.5, 1\}$ by points of different colors joined together by opaque lines. The different panels are for different orbital inclinations, with $\I/h \in \{0, 0.25, 0.5, 1\}$. Two dashed horizontal gray lines indicate, around the expected value in the limit of no gap (to the right in the plots), an error of 20\%, which is the typical uncertainty of analytical planet-disk interaction formulas \citep{2011MNRAS.410..293P}. For all inclination values, we observe a decrease in $e$-damping efficiency for deeper and deeper gaps well outside this margin of error, and down to a factor of $\sim 1/5$ less efficient eccentricity damping at the transition from type-I to type-II regimes (gap depths of $\simeq 0.3$) as compared to the limit of no gap. The data are well modelled by a double linear fit that depends on the gap depth, the eccentricity and the inclination (see Eqs.\ \eqref{eq:FitEccSegmented} and \eqref{eq:FitEccSegmented.Cont}), shown with dashed lines whose color reflects the orbital eccentricity (the slopes for the piece-wise fits are given in the legend in the top left corner of each panel).
    }
    \label{fig:edampVsObsGapDepthFitSegmented2}%
    \end{figure*}

\subsection{Eccentricity damping efficiency for partial-gap opening planets}
We show in Figure \ref{fig:edampVsObsGapDepthFitSegmented2} the observed eccentricity damping efficiency $1/\tau_e$ normalised by the expected efficiency $1/\tau_{e, \mathrm{CN2008}}$ from \cite{2008A&A...482..677C} (Eq.\ \eqref{eq:taue_CN2008}, which should be valid in the limit of no gap) as a function of the observed gap depth carved by the planet, and for different inclinations. Similarly to the 2D case, the data follow the following trends.
For low eccentricities, $0<e/h\lesssim0.5$, one can fit $1/\tau_e$ as a function of the observed gap depth (in the limit of circular/non-inclined orbits) with a straight line over the full gap depth range of interest ($\Sigma_{\min}/\Sigma_0 \simeq 0.3$ to 1). In the limit of no gap ($\Sigma_{\min}/\Sigma_0\simeq 1$), we recover the damping efficiency predicted by \cite{2008A&A...482..677C} (itself based on the results of \citealt{2004ApJ...602..388T}), as expected.\footnote{We note that in our 2D simulations the observed eccentricity damping efficiency was slightly higher than that predicted by \cite{2008A&A...482..677C}. This small inconsistency was thus due to the difference between 2D and 3D simulations as we argued in \cite{2023A&A...670A.148P}.}
For higher eccentricities, $e/h\simeq 1$, the data follow again a straight line for gap depths $\simeq 0.3$ up to $\simeq 0.8$, after which $e$-damping becomes significantly more efficient. This qualitative behaviour was already observed in \cite{2010A&A...523A..30B,2014MNRAS.437...96F} and in our 2D simulations \cite{2023A&A...670A.148P}. The reason is that shallower gaps are also thinner in radial extent, and for sufficiently high $e$, the planet's excursions around $r_\pln=a$ start interacting with the edge of the gap. Since the gap around a planet is carved where the Lindblad torques accumulate, around $r_\pln\pm 2/3 H$ \citep{2008EAS....29..165M}, this happens when $e/h\simeq 1$. The specific linear dependence of the eccentricity damping efficiency as a function of the gap depth is also observed to depend on the orbit's inclination. Given the qualitative similarities with \cite{2023A&A...670A.148P}, we obtain a fit to the data with a similar double-linear functional form which quantitatively reproduces the results of high-resolution 3D simulations:
\begin{equation}\label{eq:FitEccSegmented}
\begin{split}
    \frac{1}{\tau_e} &= \frac{1}{\tau_{e, \mathrm{CN2008}}}\times \begin{cases} 
    c_{1,e} - m_{1,e} \left(1 -\frac{\Sigma_{\min}}{\Sigma_0}\right)  & \text{if $0.3 \lesssim \frac{\Sigma_{\min}}{\Sigma_0} < 0.8$}, \\
    c_{2,e} - m_{2,e} \left(1 -\frac{\Sigma_{\min}}{\Sigma_0}\right) & \text{if $0.8 < \frac{\Sigma_{\min}}{\Sigma_0} < 1$,}
    \end{cases}
\end{split}
\end{equation}
where 
\begin{equation}\label{eq:FitEccSegmented.Cont}
\begin{split}
    m_{1,e}(e,\I)   &= 1.2 + 0.79 (e/h) (\I/h) - 0.08 (\I/h)^2,\\
    c_{1,e}(e,\I)   &= 1  + 0.71 (e/h) (\I/h) - 0.13 (\I/h)^2,\\
    m_{2,e}(e,\I)   &= \max\left\{m_{1,e}, \tilde{m}_{2,e}\right\},\\
    c_{2,e}(e,\I)   &= c_{1,e} + 0.2(m_{2,e}-m_{1,e}),\\
    \tilde{m}_{2,e}(e,\I)   &= 6.49 (e/h-0.25) + 1.61 (e/h-0.25)^2 
							-1.92 (e/h-0.25) (\I/h) + 5.14 (\I/h)^2
\end{split}
\end{equation}
Figure \ref{fig:edampVsObsGapDepthFitSegmented2} also shows a comparison between the fit (colored dashed lines) and the data (colored markers). This fit was obtained using the \texttt{LinearModelFit} function of the software package Mathematica to extract the linear models across different values of $e$ and $\I$ as a function of the gap depth, and the \texttt{NonlinearModelFit} function to extract an explicit dependence on $e$ and $\I$.\footnote{
The functional form used for the \texttt{NonlinearModelFit} function was informed by the one from \cite{2023A&A...670A.148P}, and we allowed terms up to order two in $e/h$ and $\I/h$. We investigated a fit where linear coupling terms are set to zero (as such terms are not present in \citealt{2008A&A...482..677C}) and one where such terms are allowed, and we found no significant difference in the errors for the fits. The constant term in the coefficient $c_{1,e}$ was fixed to be 1 in order to formally reproduce well known 3D damping efficiencies in the limit of no gap and vanishing eccentricities and inclinations \citep{2004ApJ...602..388T,2008A&A...482..677C}, since our data are consistent with them. The turnover value for $e/h=0.25$ found in the coefficient $\tilde{m}_{2,e}$ is similar to the one found in our previous study \cite{2023A&A...670A.148P}, but was kept as a free parameter for the fit. 
} The typical relative error given by the fit is of the order 10 -- 20\% across all eccentricity and inclination values, which is of the order of the accuracy of torque formulas from the literature. We note that the slope of the fit in the vanishing eccentricity and inclination case is the same as our 2D study \cite{2023A&A...670A.148P}.
When $\Sigma_{\min}/\Sigma_0=1$ (no gap opened by the planet), the fit recovers known 3D eccentricity damping efficiencies exactly \citep{2004ApJ...602..388T,2008A&A...482..677C}.

   \begin{figure*}
   \centering
   \includegraphics[width = 0.9\textwidth]{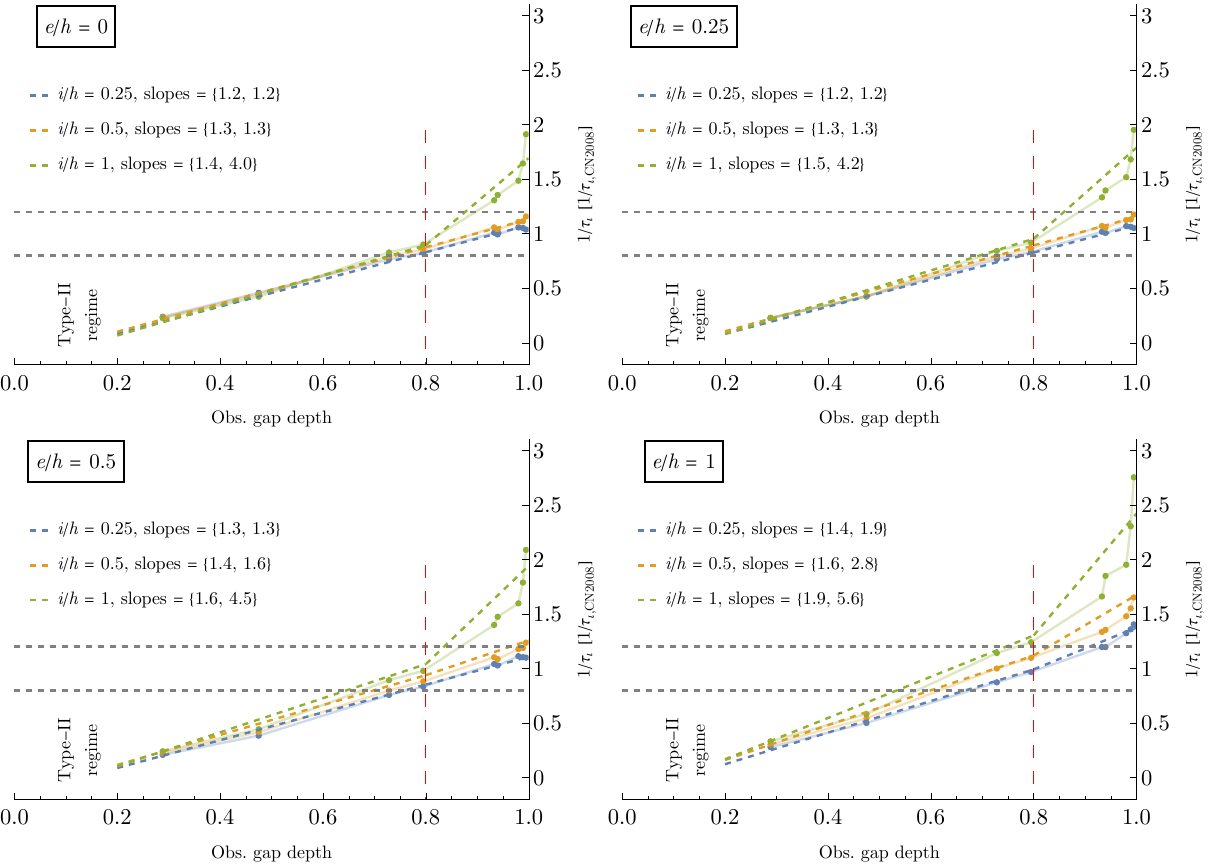} 
    \caption{
    Similar to Figure \ref{fig:edampVsObsGapDepthFitSegmented2}, but showing the observed inclination damping efficiency (normalised by the expected value from \citealt{2008A&A...482..677C}) versus the observed gap depth. Values for the $\I$-damping efficiency are shown for different inclinations $\I/h \in \{0.25, 0.5, 1\}$ by points of different colors joined together by opaque lines. The different panels are for different eccentricities, with $e/h \in \{0, 0.25, 0.5, 1\}$. Two dashed horizontal gray lines indicate, around the expected value in the limit of no gap (to the right in the plots), an error of 20\%, which is the typical uncertainty of analytical planet-disk interaction formulas \citep{2011MNRAS.410..293P}. In all panels, there is a significant decrease in $\I$-damping efficiency for deeper and deeper gaps, down to a factor of $\sim 1/5$ less efficient damping at the transition from type-I to type-II regimes (gap depths of $\simeq 0.3$) as compared to the limit of no gap. The data are well modelled by a double linear fit that depends on the gap depth, the eccentricity and the inclination (see Eqs.\ \eqref{eq:FitIncSegmented} and \eqref{eq:FitIncSegmented.Cont}), shown with dashed lines of different colors depending on the orbital inclination.
    }
    \label{fig:incdampVsObsGapDepthFitSegmented2}%
    \end{figure*}

\subsection{Inclination damping efficiency for partial-gap opening planets}
Repeating the same study for the inclination damping, we find that all the same arguments apply. Figure \ref{fig:incdampVsObsGapDepthFitSegmented2} shows the observed inclination damping efficiency $1/\tau_\I$ normalised by the expected efficiency $1/\tau_{\I, \mathrm{CN2008}}$ from \cite{2008A&A...482..677C} (Eq.\ \eqref{eq:tauinc_CN2008}), as a function of the observed gap depth carved by the planet, and for different eccentricities. The overall trends are the same as in the case of the eccentricity damping, so we obtain a fit to the data with a similar functional form:
\begin{equation}\label{eq:FitIncSegmented}
\begin{split}
    \frac{1}{\tau_\I} &= \frac{1}{\tau_{\I, \mathrm{CN2008}}}\times \begin{cases} 
    c_{1,\I} - m_{1,\I} \left(1 -\frac{\Sigma_{\min}}{\Sigma_0}\right)  & \text{if $0.3 \lesssim \frac{\Sigma_{\min}}{\Sigma_0} < 0.8$}, \\
    c_{2,\I} - m_{2,\I} \left(1 -\frac{\Sigma_{\min}}{\Sigma_0}\right) & \text{if $0.8 < \frac{\Sigma_{\min}}{\Sigma_0} < 1$,}
    \end{cases}
\end{split}
\end{equation}
where 
\begin{equation}\label{eq:FitIncSegmented.Cont}
\begin{split}
    m_{1,\I}(e,\I)      &= 1.2 - 0.19 (e/h) + 0.29 (e/h)^2 
                        + 0.18 (\I/h) + 0.41(e/h)(\I/h),\\
    c_{1,\I}(e,\I)      &= 1 - 0.19 (e/h) + 0.29 (e/h)^2 
                        + 0.36 (\I/h) + 0.41(e/h)(\I/h) -0.19(\I/h)^2 ,\\
    m_{2,\I}(e,\I)      &= \max\left\{m_{1,\I}, \tilde{m}_{2,\I}\right\},\\
    c_{2,\I}(e,\I)      &= c_{1,\I} + 0.2(m_{2,\I}-m_{1,\I}),\\
    \tilde{m}_{2,\I}(e,\I)   &= 0.8 (e/h) +1.12 (e/h)^2 + 3.14(\I/h-0.25) 
                            -0.42 (e/h) (\I/h-0.25) + 2.9 (\I/h-0.25)^2.
\end{split}
\end{equation}
This double-linear fit is shown in Figure \ref{fig:incdampVsObsGapDepthFitSegmented2}. The typical error given by the fit is less than 10\% across all inclination and eccentricity values, and for $\Sigma_{\min}/\Sigma_0=1$ (no gap opened by the planet), we recover known 3D eccentricity damping efficiencies \citep{2004ApJ...602..388T,2008A&A...482..677C}. We note that, unlike in \eqref{eq:FitEccSegmented.Cont}, there are coupling terms that are linear in e; this is in contrast with \cite{2008A&A...482..677C}'s fit, which has no such terms. We checked that imposing that there be no linear coupling terms yields a worse fit to the data. Therefore, although we do not have a physical explanation for these terms, we use an agnostic approach and keep them in the fit for $\tau_\I$ to ensure the best possible match with the outcome of hydrodynamical simulations.

\section{Discussion}\label{sec:Discussion}

\subsection{Pebble isolation mass}
In our simulations, we set the planetary mass as a free parameter. However, in a real protoplanetary disk, this mass (at least in the limit of partial-gap opening planets where gas accretion is a negligeable effect) will be the result of accretion of solid, which is itself a process that depends on the gas structure and planet-disk interaction. In the pebble accretion scenario (\citealt{2010A&A...520A..43O,2012A&A...544A..32L}; see \citealt{2017AREPS..45..359J} for a review), the maximum mass that can be reached is the so-called pebble isolation mass \citep{2012A&A...546A..18M,2014A&A...572A..35L,2018A&A...615A.110A,2018A&A...612A..30B,2018ApJ...854..153W}. This is the mass at which the planet disturbs the disk surface density enough that it creates a pressure barrier outside of its orbit which prevents further pebbles to drift inwards and be accreted onto the planet.

Our high-resolution 3D simulations can also be used to validate previous works on the pebble isolation mass \citep{2018A&A...612A..30B}.
From the output of our simulations, we define the gas density $\rho(r) = \Sigma(r)/(\sqrt{2\pi} h r)$ and the pressure $P(r) = c_\mathrm{s}^2 \rho(r)$, where $c_\mathrm{s} = h r \Omega_\Kepl$ is the sound speed. We thus consider the pressure gradient $\frac{\partial \log P}{\partial \log r}$ and check when it changes sign in the vicinity of the planet's orbit.
We then compare this outcome with the prediction from \cite{2018A&A...612A..30B} for the pebble isolation mass:
\begin{equation}\label{eq:PIM}
\begin{split}
        m_\mathrm{p.iso} &= 25 M_\oplus \times 
        \left(\frac{h}{0.05}\right)^3 \left(0.34 \left(\frac{\log_{10}(10^{-3})}{\log_{10}(\alpha)}\right)^4 + 0.66\right)\left(1-\frac{\frac{\partial \log P}{\partial \log r}+2.5}{6}\right).
\end{split}
\end{equation}
We find that this prediction aligns well when our results, within an uncertainty of 20\%, across all values of aspect ratios and viscosities considered here: when a planet is predicted to be below the pebble isolation mass, it does not generate a pressure barrier outside of its orbit in our simulation, and, conversely, when a planet is predicted to be above the pebble isolation mass, it generates a strong pressure bump. Note that \cite{2018A&A...612A..30B} also used 3D simulations run with the \texttt{fargOCA} code, albeit with a lower resolution. 
Thus, the prescriptions in Equations \eqref{eq:FitEccSegmented}, \eqref{eq:FitEccSegmented.Cont}, \eqref{eq:FitIncSegmented}, \eqref{eq:FitIncSegmented.Cont} for the orbital damping timescales, together with \cite{2018A&A...612A..30B}'s formula \eqref{eq:PIM} for the pebble isolation mass give a complete analytical description of the evolution of a super-Earth/Mini-Neptune planet embedded in a disk, from its limiting mass growth to its dynamical response to the presence of the disk, which is consistent with high-resolution 3D locally isothermal hydrodynamical simulations.

\subsection{Application to $N$-body integrations}

\begin{figure}[t!]
\centering
\includegraphics[width= 0.24\textwidth]{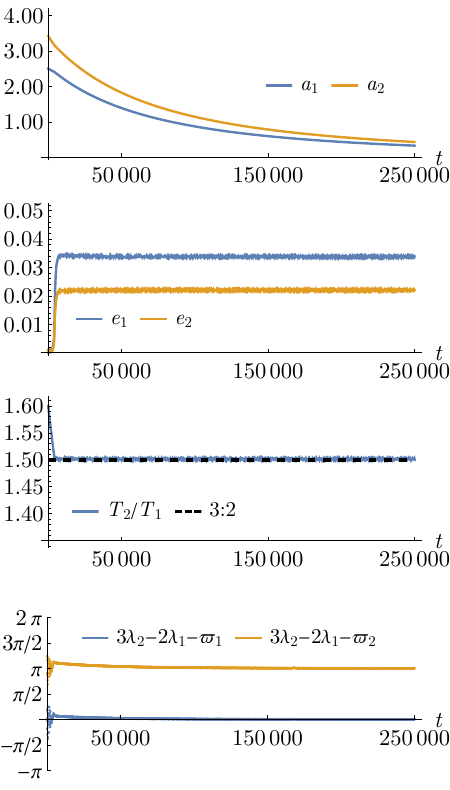}%
\includegraphics[width= 0.24\textwidth]{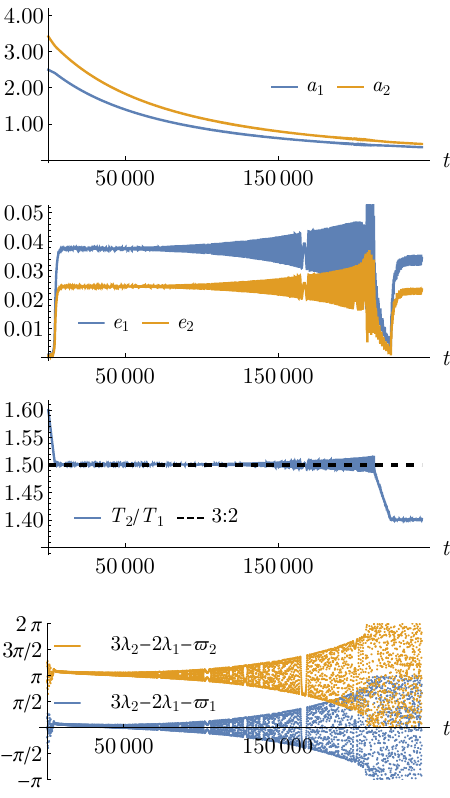}
    \caption{
    Evolution under convergent migration of two planets, with masses $m_1 = 3\times 10^{-5} M_*$ and $m_2 = 5\times 10^{-5} M_*$ respectively, near the 3:2 commensurability. The disk has $\alphaturb=3.16\times10^{-4}$ and $h=0.05$. Panels on the left and on the right have the same initial conditions, and show, from top to bottom, the evolution of the semi-major axes, eccentricity, period ratio and resonant angles. All panels make use of \cite{2011MNRAS.410..293P}'s torque prescription; instead, panels on the left use \cite{2008A&A...482..677C}'s $e$-damping prescription (eq.\ \eqref{eq:taue_CN2008}), while panels on the right use our modified $e$-damping prescription (eq.\ \eqref{eq:FitEccSegmented}). The modified damping efficiency manifests itself in that, due to the overall less efficient $e$-damping, the captured resonant state is found at higher eccentricities; moreover, the libration inside the 3:2 resonance becomes overstable in the right panels, leading to an escape from the resonance and a more compact final state.
    }
\label{fig:3-2}%
\end{figure}

\begin{figure}[t!]
\centering
\includegraphics[width= 0.24\textwidth]{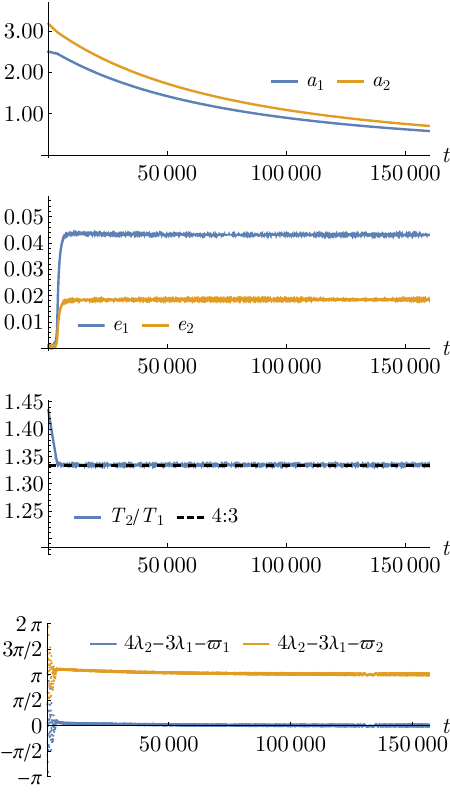}%
\includegraphics[width= 0.24\textwidth]{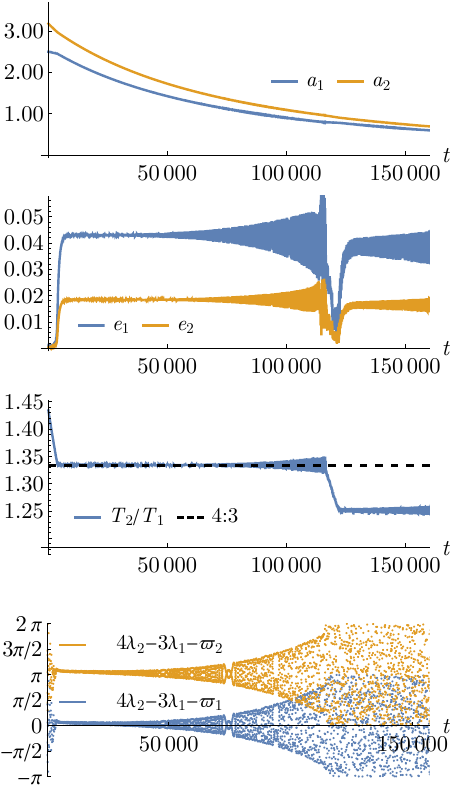}
    \caption{
    Similar to Figure \ref{fig:3-2}, but for two planets with masses $m_1 = 2\times 10^{-5} M_*$ and $m_2 = 5\times 10^{-5} M_*$ in the vicinity of the 4:3 mean motion resonance.
    }
\label{fig:4-3}%
\end{figure}

Analytical formulas for planet-disk interactions are widely used in the literature in the context of planet population synthesis models \citep{2008ApJ...673..487I,2009A&A...501.1139M,2013A&A...558A.109A,
2017MNRAS.464..428A,
2017MNRAS.470.1750I, 
2018MNRAS.474..886N, 2019A&A...623A..88B, 
2021A&A...650A.152I, 2021A&A...656A..69E}. In particular, convergent migration (when two planets orbiting the same disk migrate in such a way that the sizes of their orbits approach each other) and eccentricity damping are generally associated to the assembly of mean motion resonant chains \citep{2007ApJ...654.1110T,2008A&A...482..677C,2008A&A...478..929M}. However, the specific resonances that are built crucially depend on orbital damping efficiencies. This is important not only because the resonant structure attained at the end of the disk phase will be different on a quantitative level (i.e.,\ which resonances will be observed in a given system), but also because the stability properties of these configurations are dependent on the resonance. More precisely, whether or not a given resonant chain assembled via disk-driven convergent migration will go unstable after the disappearance of the disk depends on how compact the chain is \citep{2020MNRAS.494.4950P,2022Icar..38815206G}. Thus, although migration within a disk does in general lead to the assembly of resonant chains, which resonances are built has a strong impact in whether or not these resonance will even be observable after the the removal of the gas disk.

The processes of resonant capture and which resonances will be built under which conditions are fairly well understood \citep{2015MNRAS.451.2589B,2015ApJ...810..119D,2018CeMDA.130...54P,2023ApJ...946L..11B}. A mean motion resonance can be skipped if the resonance crossing time for the two planets is comparable to or shorter than the period of the planets' resonant interaction (the so called adiabatic limit, which is a condition on the torque), or if the dissipative torque is simply stronger than the resonant torque (which is a condition on the relative disk-driven $e$-damping onto the planets) \citep{2015MNRAS.451.2589B,2023ApJ...946L..11B}. Moreover, even when the evolution is adiabatic and a resonance is successfully established, the presence of eccentricity damping breaks in general the adiabatic regime and the resonant equilibrium point may become an unstable fixed point, leading to so-called overstable librations and the escape from the resonant state \citep{2014AJ....147...32G}. For a fixed mass ratio between the planets, this is essentially a condition on the relative eccentricity damping efficiencies onto the two planets (\citealt{2015ApJ...810..119D,2018MNRAS.481.1538X}). Thus, even when the torques satisfy the adiabatic and stability conditions, different eccentricity damping efficiencies will lead to different final (resonant) states. 

To elucidate this point, we show examples of $N$-body integrations of two planets inside a disk in Figures \ref{fig:3-2} and Figures \ref{fig:4-3} for the 3:2 and 4:3 resonance, respectively. These examples are not meant to depict realistic resonant capture scenarios, but rather to stress the differences that arise from the two eccentricity damping prescriptions. For this reason, we consider a constant disk surface density and aspect ratio, so that when the planets migrate inward their planet-disk interactions remain unchanged. We take $\alphaturb=3.16\times10^{-4}$ and $h=0.05$.
In Figure \ref{fig:3-2}, we simulate two planets with masses $m_1/M_* = 3\times 10^{-5}$ and $m_2/M_* = 5\times 10^{-5}$ (which are below the pebble isolation mass for these disc parameters), starting with initially circular and coplanar orbits, and with initial period ratio slightly larger than 3:2. Since the surface density is constant at all radii, and planet 2 is more massive than planet 1, it will migrate inward faster than planet 1, so their period ratio will decrease and the two planets will approach the 3:2 commensurability under convergent migration. What happens after this is different in the panels on the right compared to the ones on the left. 
In the panels on the left, we implemented \cite{2011MNRAS.410..293P}'s torque prescription in conjunction with \cite{2008A&A...482..677C}'s eccentricity damping formula, as is commonly done in the literature (e.g.\ \citealt{2017MNRAS.470.1750I, 2021A&A...650A.152I, 2021A&A...656A..69E}). We observe that a successful capture has occurred (the evolution is in the adiabatic limit) and the resonant state achieved is stable (no overstable libration and jumping out of the resonance after capture). 
In the panels on the right, we used exactly the same initial conditions but we implemented our modified eccentricity damping formula \eqref{eq:FitEccSegmented}. The planets have estimated gap depths of 0.87 and 0.67 respectively, so eccentricity damping is less efficient overall and the planets attain higher eccentricities ($e$-damping on planet 1 is very similar under both prescription, while planet 2 undergoes a less efficient $e$-damping with the modified formula \eqref{eq:FitEccSegmented}, given its deeper gap). We observe that the final resonant state is overstable, that is, the amplitude of libration increases in time, and the system jumps out of the 3:2 resonance, to eventually end up in a more compact configuration.
Figure \ref{fig:4-3} shows a similar result in the case of the 4:3 mean motion resonance, with planet masses $m_1/M_* = 2\times 10^{-5}$ and $m_2/M_* = 5\times 10^{-5}$, where again $\alphaturb=3.16\times10^{-4}$ and $h=0.05$. This case appears even more elusive at first, as the capture eccentricities appear to be very similar, but the stability properties of the resonant equilibrium are not. This is because in this case the eccentricity damping on the inner planet (which has a gap depth of 0.94) is enhanced, while that on the outer planet (which has a gap depth of 0.67) is less efficient using our modified formula \eqref{eq:FitEccSegmented} compared to the pure \citep{2008A&A...482..677C} prescription; thus, although the total eccentricity damping onto the planets is similar, the relative $e$-damping is different, the system undergoes overstable libration and exits the resonance.

In both cases, the final state will be more compact using the modified $e$-damping prescription \eqref{eq:FitEccSegmented} than using \cite{2008A&A...482..677C}'s formula \eqref{eq:taue_CN2008}, leading to a system more susceptible to instabilities after the disk is removed \citep{2020MNRAS.494.4950P,2022Icar..38815206G}. Even if the libration does not become overstable, the final eccentricities inside the same resonance may be higher for planets opening moderate gaps, which is also known to lead to less stable systems \citep{2018CeMDA.130...54P,2020MNRAS.494.4950P}.
Although simple by design, these experiments show that taking into account a more realistic modeling of orbital damping efficiencies for partial gap opening planets may have noticeable effects in the final product of population synthesis models. In particular, it may resolve the need to resort to more massive planets in order to trigger the instabilities needed to explain the orbital period distribution of known exoplanets \citep{2017MNRAS.470.1750I,2021A&A...650A.152I}.

Planetary inclination are affected by mean motion resonances only by second order effects (i.e.,\ terms that are proportional to $e\times\I^2$ for first order resonances). In population synthesis models, they typically arise from seeding the planets with small initial inclinations, which may be subsequently excited by close encounters, collisions and scattering events. Because of the stochastic nature of this process, we do not systematically investigate the details of how our modified damping formulas \eqref{eq:FitIncSegmented}, \eqref{eq:FitIncSegmented.Cont} might impact population synthesis calculations. In general, we expect that mutual inclinations will be enhanced, especially for planets opening deep gaps, because of the reduced $\I$-damping efficiency in these cases.

\section{Conclusions}\label{sec:Conclusion}
In this paper we investigated planet-disk interactions for partial gap opening planets using high-resolution 3D hydro-dynamical simulations of locally isothermal disks with varying levels of turbulent viscosities and aspect ratios with an embedded planet of varying mass. The goals and methodology are similar to the ones used in a previous paper \citep{2023A&A...670A.148P}, which was limited to the 2D case: we reconsidered the problem of orbital damping timescales for planets that are classically in the type-I migration regime (a few to a few ten's of Earth's mass) but which would open partial gaps in disks of low viscosity and/or in thin disks, and are thus in between the type-I and type-II regimes. The expression of the torque felt by such planets in disks of arbitrary viscosities has been the subject of various works (e.g.\ \citealt{2006Icar..181..587C, 2011MNRAS.410..293P, 2017MNRAS.471.4917J, 2018ApJ...861..140K}) and these migration prescriptions have been used in population synthesis models to reproduce the observed characteristics of exoplanetary systems (e.g.\ \citealt{2017MNRAS.470.1750I, 2018MNRAS.474..886N, 2019A&A...623A..88B, 2020ApJ...892..124O, 2021A&A...650A.152I, 2021A&A...656A..69E}). The transition between classical type-I and type-II torques for partial gap opening planets on circular and non-inclined orbits has been shown to depend linearly on the gap depth carved by the planet \citep{2018ApJ...861..140K}, and we showed in \cite{2023A&A...670A.148P} that an equivalent linear trend is also observed in the eccentricity damping efficiency. In particular, $e$-damping efficiencies can be significantly lower (i.e.,\ eccentricity damping timescales can be longer) than in the case of shallow gaps, which may have important consequences on the outcome of population synthesis models. This also fills the gap between the observed eccentricity damping that is typically associated to low-mass planets and the eccentricity pumping that is observed for very high mass planets \citep{2001A&A...366..263P,2006A&A...447..369K,2013A&A...555A.124B}. 

Here, we extend the study to 3D disks and allow the planets to reside on orbits that are eccentric as well as inclined with respect to the disk mid-plane. We considered Super-Earth-type planets of varying fixed masses ($\mpl/M_* = 1$ to $6 \times 10^{-5}$) and varying orbital eccentricities and inclinations ($e/h$ and $\I/h$ ranging from $0$ to 1, where $h=H/r$ is the disk's aspect ratio), embedded in disks of varying viscosities ($3.16\times10^{-5}$ to $\alphaturb = 10^{-3}$) and aspect ratios ($h = 0.04$ to $0.06$). The planet is kept on a fixed orbit and the system is evolved for thousands of the planet's orbital period until a steady-state is achieved.

We analysed the surface density profile of the disk in response to the presence of the planet, in particular the depth of the gap opened by the planet and the threshold beyond which vortices appear, which gave similar results to the 2D case \citep{2023A&A...670A.148P}. Our fit for the gap depth agrees well with the one from \cite{2018ApJ...861..140K} in the higher-viscosity disks, but it better reproduces the observed gap in the low-viscosity regime. We also considered the establishment of a pressure bump outside the orbit of the planet that would cause the inflow of pebbles to stop, thus halting the accretion of solid material; we found that our simulations agree well with previous results on the scaling of the pebble isolation mass with planetary and disk parameters \citep{2018A&A...612A..30B}.

We then considered the eccentricity and inclination damping efficiencies and their dependence on the gap depth. We found a similar qualitative behaviour as in our 2D study \cite{2023A&A...670A.148P}. The orbital damping efficiencies (rescaled by the expected efficiencies in the no-gap case, \citealt{2004ApJ...602..388T,2008A&A...482..677C}) are well described as linear functions of the gap depth with slopes and intercepts that depend in general on the eccentricity and inclination; a break is observed around gap depths of 80\% after which, for shallower gaps, the damping efficiency's slope with respect to the gap depth increases (see Fig.'s \ref{fig:edampVsObsGapDepthFitSegmented2} and \ref{fig:incdampVsObsGapDepthFitSegmented2}). These features can be understood on theoretical grounds \citep{2023A&A...670A.148P}. We therefore used an equivalent functional form as our 2D fit from \cite{2023A&A...670A.148P} and obtained an explicit but simple formula that depends on the gap depth, the orbital eccentricity and inclination (Eq.'s \eqref{eq:FitEccSegmented}, \eqref{eq:FitEccSegmented.Cont} and \eqref{eq:FitIncSegmented}, \eqref{eq:FitIncSegmented.Cont}), and which approximates the outcome of 3D high-resolution hydrodynamical simulations within the errors of torque formulas commonly used in population synthesis works. This gives a simple but complete description of planet-disk interactions for partial gap opening planets (from the traditional type-I regime down to gaps close to the traditional type-II regime) to be used in $N$-body simulations.

Finally, we tested the consequences of our novel formulas in the context of planet population synthesis simulations, in particular in the formation of mean motion resonant chains. Resonances are naturally associated with convergent migration (when two planets orbit the same star embedded in the same protoplanetary disk and the sizes of their orbits change in such a way that the orbits get closer to each other,  \citealt{2007ApJ...654.1110T,2008A&A...482..677C,2008A&A...478..929M}). This process is well understood on theoretical grounds (e.g.,\ \citealt{2013A&A...556A..28B,2015MNRAS.451.2589B}), and in particular it is known that which mean motion resonances are skipped and which ones are successfully established depends on the orbital damping timescales \citep{2015MNRAS.451.2589B,2015ApJ...810..119D,2018MNRAS.481.1538X,2023ApJ...946L..11B}, and so do the final eccentricities after successfully capturing in a given mean motion resonance \citep{2005MNRAS.363..153P,2008A&A...483..325C,2014AJ....147...32G,2015ApJ...810..119D,2018CeMDA.130...54P}. We show simple examples in which our modified formulas for orbital damping timescales for partial-gap opening planets yield dynamically different results than the prescriptions used so far in the literature, and we stress in what way this would impact the orbital states obtained at the end of population synthesis models. In particular, the establishment of more dynamically excited and compact states may resolve the necessity for more massive planets in order to trigger the instabilities that can explain the orbital period distribution of known exoplanets \citep{2017MNRAS.470.1750I,2021A&A...650A.152I}.

\section{Acknowledgments}
The authors are grateful to the anonymous referee for comments which improved the clarity and content of the manuscript. G.\ P.\ and B.\ B.\ thank the European Research Council (ERC Starting Grant 757448-PAMDORA) for their financial support. G.\ P.\ also thanks the Barr Foundation for their financial support, and K.\ Batygin for helpful comments that improved the manuscript.
E.\ L.\ whishes to thank Alain Miniussi for the maintainance and re-factorization of the code fargOCA.
We acknolewdge HPC resources from GENCI DARI n. A0140407233.

%



\software{fargOCA (\url{https://gitlab.oca.eu/DISC/fargOCA}), Mathematica (\url{https://www.wolfram.com/mathematica}), swift (\url{https://www.boulder.swri.edu/~hal/swift.html}).}



\appendix

\section{Fixed vs.\ free planets}\label{apx:FreePlanets}
In this section we compare the eccentricity and inclination damping timescales that one would infer by i) keeping the planet on a fixed orbit, or ii) letting the planet respond to the disk and fitting the evolution of the osculating orbital elements. The first method is the one we used in this paper, and which is also typically used to obtain the strength of the torque \citep{2011MNRAS.410..293P,2017MNRAS.471.4917J}; the advantage of this method is that one can wait arbitrarily long until a steady state is reached without having to worry about the planet's orbital state changing over time. The second method is the one used e.g.\ in \cite{2008A&A...482..677C} and links more directly to practical applications for population synthesis works and $N$-body integrations.

To check if the two methods agree, we run additional hydro-dynamical simulations where we track the time evolution of the orbital elements after releasing a planet on an initially eccentric and/or inclined orbit. In particular, we use as initial conditions (for both the gas and the planet) the end state of our simulations where the planet had been kept on a fixed orbit, after a steady-state has been reached, and we release the planet. We then set up $N$-body integrations mimicking planet-disk interactions with the same initial conditions as the hydro-simulations and check whether the evolutions of the eccentricity/inclination over time match with that of the hydro-simulations, that is, whether they are damped on a timescale comparable to the expected one. In these $N$-body simulations, we use both the classical \cite{2008A&A...482..677C}, and our modified prescription from Eq.'s \eqref{eq:FitEccSegmented}, \eqref{eq:FitEccSegmented.Cont} and \eqref{eq:FitIncSegmented}, \eqref{eq:FitIncSegmented.Cont} to mimic disk-driven $e$- and $\I$-damping.

\begin{figure*}
   \centering
   \includegraphics[width = 0.275\textwidth]{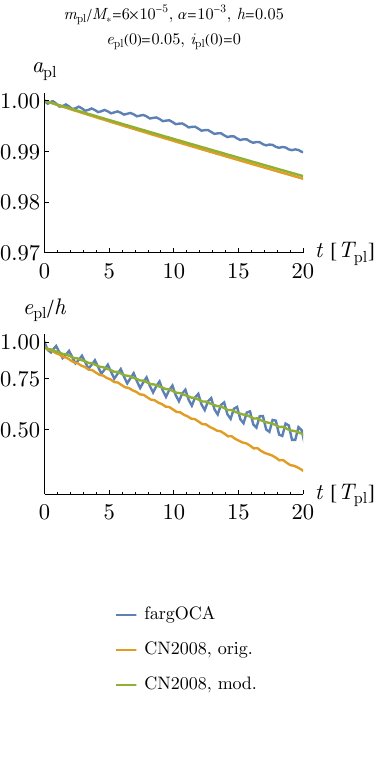}\quad
   \includegraphics[width = 0.275\textwidth]{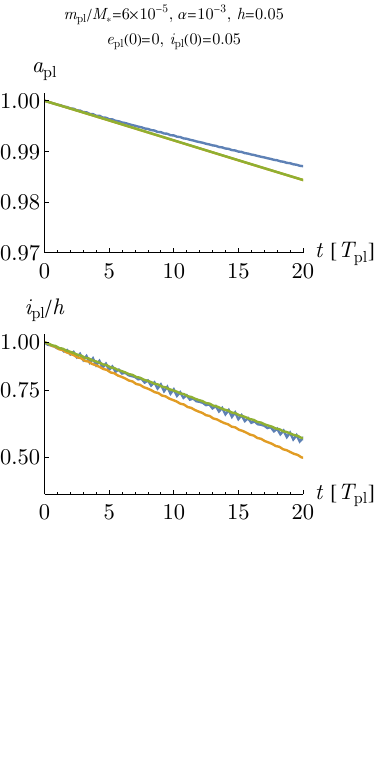}\quad
   \includegraphics[width = 0.275\textwidth]{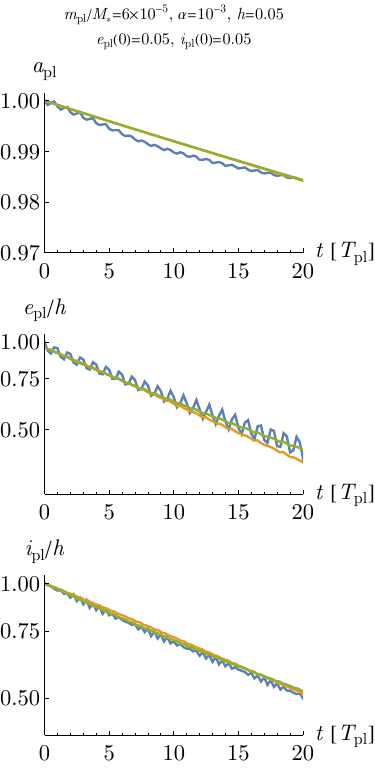}
    \caption{Evolution over time (in units of the orbital period $T_\mathrm{pl}$ at $a=a(0)=1$) of the orbital elements for a planet subject to disc-driven migration and $e$- and $\I$-damping. The planet and disk parameters are reported above each panel. In all panels we plot the outcome of a hydro-simulation (labeled ``fargOCA'') and of two $N$-body integrations (one with the classical \cite{2008A&A...482..677C} damping prescription, labeled ``CN2008, orig.'', and one with our modified prescription for $\tau_e$ and $\tau_\I$, labeled ``CN2008, mod.''), as shown in the legend at the bottom. Each column represents a different initial condition for the eccentricity and inclination of the planet, where either $e$, or $\I$, or both, are initialised at $h$. We plot on the top row the evolution of the semi-major axis (notice that the curves for the two $N$-body integrations largely overlap) and on the bottom rows that of the eccentricity and/or inclination. Notice the log scale on the vertical axis, which we use in order to observe the slope of the curves, which is a measure of $\tau_e$ and $\tau_\I$; the small oscillations in orbital elements (noticeable especially in the eccentricity evolution) have a frequency equal to the orbital frequency and have a relatively constant amplitude.
    }
    \label{fig:mpl-6.e-5_alpha-1.e-3_h-0.05_free-planet}%
\end{figure*}

\begin{figure*}
   \centering
   \includegraphics[width = 0.275\textwidth]{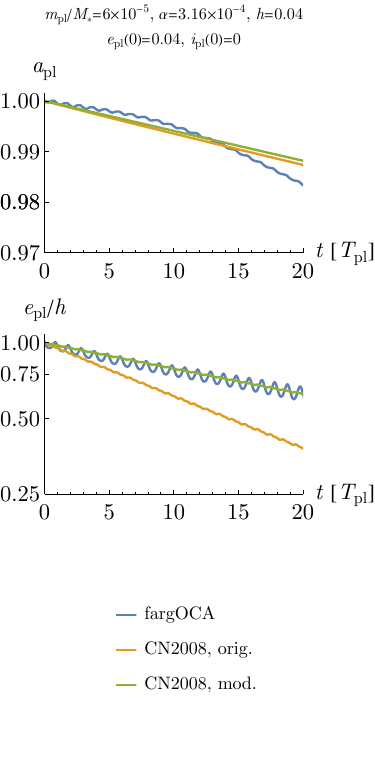}\quad
   \includegraphics[width = 0.275\textwidth]{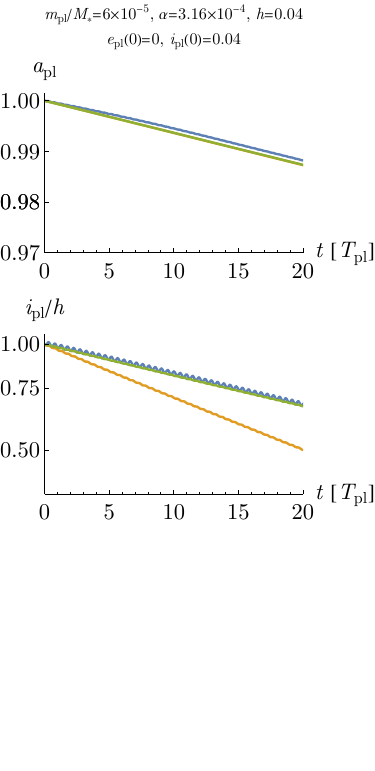}\quad
   \includegraphics[width = 0.275\textwidth]{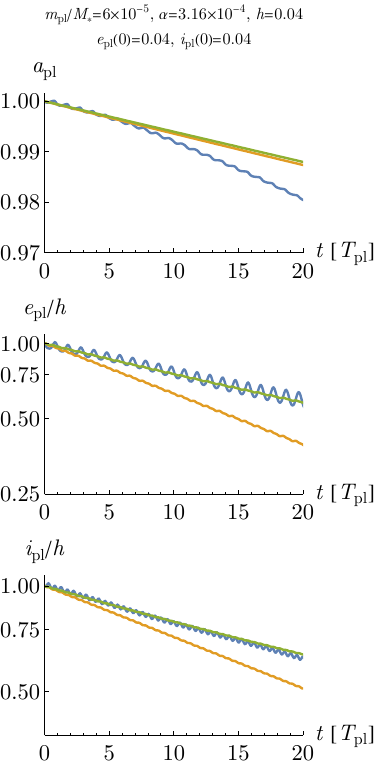}
    \caption{Similar to Fig.\ \ref{fig:mpl-6.e-5_alpha-1.e-3_h-0.05_free-planet}, but for planet and disk parameters such that the gap carved by the planet is deep enough that we would expect noticeable $e$- and $\I$-damping efficiencies between the classical \cite{2008A&A...482..677C} prescription and our modified damping formulas. Indeed, we observe less efficient damping in all cases, with our prescription following very closely the eccentricity and inclination evolution of the hydro-dynamical simulations. 
    }
    \label{fig:mpl-6.e-5_alpha-3.16e-4_h-0.04-hat-e_free-planet}%
\end{figure*}

Since the typical type-I damping timescale $\tau_\mathrm{wave}$ is inversely proportional to $m_\mathrm{pl}$ (Eq.\ \eqref{eq:tauwave}), in order to spare computational resources we consider the cases with the highest planetary mass $m_\mathrm{pl}/M_*=6\times10^{-5}$. We then first consider disk parameters such that our fitting formulas do not constitute a significant deviation from \cite{2008A&A...482..677C} (e.g.\ $h=0.05$, $\alphaturb=10^{-3}$) in order to make a fair comparison. Secondly, we consider one case where we instead expect there to be a noticeable difference in damping timescales (e.g.\  $h=0.04$, $\alphaturb=3.16\times10^{-4}$, which gives an estimated gap depth of $\simeq 0.3$, and thus a factor $\sim 4$ difference in damping efficiencies). We consider as initial conditions $e/h$ and/or $\I/h$ equal to 1, and we run the hydrodynamical simulations for 20 orbits, which is enough to track the damping in the eccentricities/inclinations.
The results of the $m_\mathrm{pl}/M_*=6\times10^{-5}$, $h=0.05$, $\alphaturb=10^{-3}$ setup are presented in Figure \ref{fig:mpl-6.e-5_alpha-1.e-3_h-0.05_free-planet}, which shows the evolution of the semi-major axis and eccentricity and/or inclination for a planet with different combinations of initial conditions (eccentric but coplanar with the disk, inclined but circular, eccentric and inclined). In this case, as expected, the outcomes of $N$-body integrations with the standard \cite{2008A&A...482..677C} prescription (labeled ``CN2008, orig.'') and with the modified prescription from this paper (labeled ``CN2008, mod.'') are rather similar to each other\footnote{
We note that the label ``CN2008'' refers to \cite{2008A&A...482..677C}'s prescription for $\tau_e$ and $\tau_\I$, while $\tau_\mig$ (i.e.\ the torque) is in both cases obtained from \cite{2011MNRAS.410..293P}'s prescription, modulated by the gap depth as proposed in \cite{2018ApJ...861..140K}, as well as by factors dependent on $e$ and $\I$ (\citealt{2013A&A...553L...2C, 2013A&A...558A.105P, 2014MNRAS.437...96F}; see also Sect.\ \ref{subsec:TIForces}). This is what is typically done in population synthesis works, and we use it here so that the evolution of the semi-major axis (which should be considered as irrelevant a factor as possible for our scope here) is as close as possible to the outcome of the hydro-dynamical simulations.}. They also match well against the outcome of the hydro-dynamical simulations with the released planet. When \cite{2008A&A...482..677C}'s and our prescription differ more significantly (e.g., in the left and middle columns), the latter gives a better match to the outcome of our hydro-dynamical simulations.
In a similar fashon, the results of the $m_\mathrm{pl}/M_*=6\times10^{-5}$, $h=0.04$, $\alphaturb=3.16\times10^{-4}$ setup are shown in Figure \ref{fig:mpl-6.e-5_alpha-3.16e-4_h-0.04-hat-e_free-planet}. Here, the difference in evolution for the $N$-body integrations with the different damping prescriptions is more apparent as expected. We see that our modified prescription which takes into account the partial gap opened by the planet yields a good match to the outcome of these hydro-dynamical simulations. This analysis also shows that, within the parameter space considered in this work, whether a planet is kept on a fixed orbit or whether it is allowed to move in response to the gas would result in comparable orbital damping timescales.

\section{Convergence tests}\label{apx:Convergence}

   \begin{figure*}
   \centering
   \includegraphics[width = 0.45\textwidth]{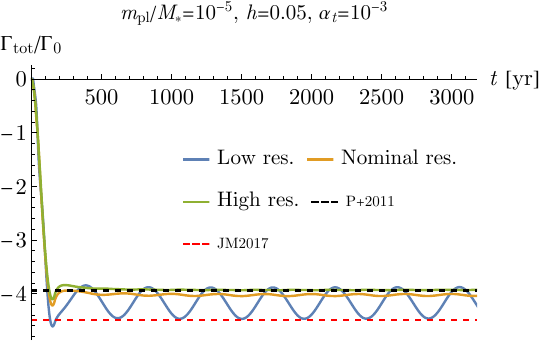}%
   \includegraphics[width = 0.45\textwidth]{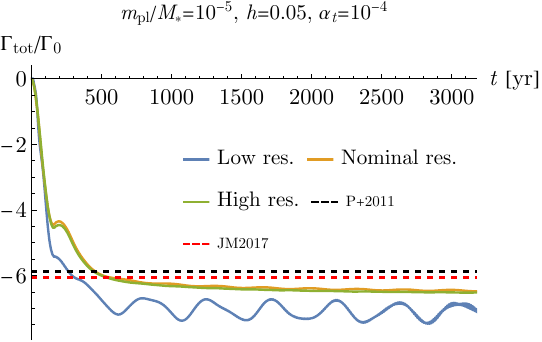}%
    \caption{Resolution convergence test in two setups with $\alphaturb=10^{-3}$ (left panel) and $\alphaturb=10^{-4}$ (right panel); in both panels, $m_\pln/M_* = 10^{-5}$ and $h=0.05$. Continuous colored lines represent the torque output of different 3D simulations with different resolutions as shown in the legend. The torques are rescaled by the torque factor $\Gamma_0:=(m_\pln/M_*)^2 \Sigma r^4 \Omega_\Kepl^2/h^2$. Dashed lines show the prediction from \cite{2011MNRAS.410..293P}'s torque prescription (P+2011) and \cite{2017MNRAS.471.4917J}'s torque prescription (JM2017).
    }
    \label{fig:Convergence}%
    \end{figure*}

We run a few resolution tests to check the convergence of the results of our hydro-dynamical simulations. In all cases, the azimuthal extent is the full $[0,2\pi)$ interval and the colatitude is $83^\circ$.\\

Res.\ L). Radial extent from $0.35$ to $3$ AU (like our 2D paper), $N_r=512$, $N_\phi=1500$ and $N_\theta=64$, $\delta_r \simeq \delta_\phi \simeq \delta \theta \simeq 0.005$.\\

Res.\ N). Radial extent from $0.5$ to $2$ AU, $N_r=512$, $N_\phi=2000$ and $N_\theta=70$, $\delta_r \simeq \delta_\phi \simeq \delta\theta \simeq 0.003$.\\

Res.\ H). Radial extent from $0.5$ to $2$ AU (like our Res.\ N), $N_r=600$, $N_\phi=2500$ and $N_\theta=100$, $\delta_r \simeq \delta_\phi \simeq \delta\theta \simeq 0.0025$.\\

All three gave similar results. At low $\alpha\sim 10^{-4}$, the lower resolution run appears slightly different from the higher resolution ones in that the disk goes mildly unstable, while Res.\ N and Res.\ H were still extremely similar. We thus use Res.\ N as our nominal resolution in order to spare computational resources while maintaining a good accuracy in our results. Figure \ref{fig:Convergence} shows two examples of our resolution tests.


\bibliography{PBL_3D}{}
\bibliographystyle{aasjournal}



\end{document}